\documentclass[preprints,review,accept,moreauthors,a4paper]{mdpi}
\firstpage{1}
\pubvolume{xx}
\issuenum{1}
\articlenumber{5}
\pubyear{2020}
\copyrightyear{2020}
\usepackage{amssymb}
\usepackage{graphics}
\usepackage{amsmath}
\usepackage{amsfonts}
\usepackage{color}

\definecolor{darkblue}{rgb}{0,0,.7}

\history{Received: date; Accepted: date; Published: date}

\Title{Dark side of Weyl gravity}

\Author{Petr Jizba $^{1,\ddagger}$*\orcidA{}, Les\l{}aw Rachwa\l{} $^{1,\ddagger}$\orcidB{}, Stefano~G. Giaccari $^{2,\ddagger}$\orcidC{}  and Jaroslav K\v{n}ap $^{1,\ddagger}$}

\AuthorNames{Petr Jizba, Les\l{}aw Rachwa\l{}, Stefano~G. Giaccari and Jaroslav K\v{n}ap}

\address{%
$^{1}$ \quad FNSPE, Czech Technical University in Prague, B\v{r}ehov\'{a} 7, 115 19 Praha 1, Czech Republic; \\
$^{2}$ \quad Department of Sciences,
Holon Institute of Technology (HIT),
52 Golomb St., Holon 5810201, Israel;
}
\corres{Correspondence: p.jizba@fjfi.cvut.cz; Tel.: +420 775-317-307 (P.J.)}
\secondnote{These authors contributed equally to this work.}

\abstract{With the help of a functional renormalization group, we study the dynamical breakdown of scale invariance in quantum Weyl gravity by starting from the UV fixed point that we assume to be Gaussian. To this end, we resort to two classes of Bach-flat backgrounds, namely maximally symmetric spacetimes and Ricci-flat backgrounds in the improved one-loop scheme. We show that apart from a genuine IR fixed point that is reached at a zero value of the running scale, the renormalization group flow also exhibits bouncing behavior. We demonstrate that the IR fixed point found is IR-stable in the space of the considered couplings. As a next step, we analyze physics in the broken phase. In particular, we show that in the low-energy sector of the broken phase, the theory looks like Starobinsky $f(R)$ gravity with a gravi-cosmological constant that has a negative sign in comparison to the usual matter-induced cosmological constant. We discuss implications for cosmic inflation and highlight a non-trivial relation between Starobinsky’s parameter and the gravi-cosmological constant. Salient issues, including the scheme independence of the IR fixed point and the role of trace anomaly, are also discussed.
}


\begin{document}


\section{Introduction}

It was just two years after the birth of the Yang--Mills (YM) gauge theory~\cite{YM} when Utiyama in his 1956 seminal paper~\cite{Utiyama}, recognized the similarity between gravity and YM fields, and started new  research line known as ``gauge theories of gravity''~\cite{Kibble,Neeman,Ivanov}. The central focus of such a program was based on gauging Lorentz, Poincar\'{e} or de-Sitter groups and this, in turn, has led to gravity theories (including Einstein--Hilbert, Einstein--Cartan theories, etc.) that  despite their formal appeal share the same fate with Einstein's general relativity. Namely, such theories have dimensionful couplings and hence they are perturbatively non-renormalizable when quantized.
%
More recently the deep relationship between gauge and gravity theories has further been explored in the light of the holographic principle (as realized, for example, by the AdS/CFT correspondence~\cite{Maldacena:1997re}) and the ``gravity=gauge $\times$gauge''  principle  (as embodied, for example, in the Ben--Carrasco--Johansson color-kinematic correspondence~\cite{Borsten:2020bgv}).

Both Lorentz, Poincar\'{e} and de-Sitter groups are subgroups of the $15$-parameter conformal group $O(4,2)$, which is
the group of spacetime transformations that leave invariant the null interval.
It is known that YM theories have a number of desirable features when quantized and this, in part, comes from their conformal  invariance. Indeed, the conformal invariance of the YM action immediately implies that no massive (or dimensionful) coupling is present and so the theory is  (power-counting) renormalizable. Conformal invariance  is also instrumental in finding self-dual YM instantons~\cite{Taubes1,Taubes2}.
%
%
One might thus expect that by gauging the conformal group one obtains gravity theory that would inherit on a quantum level some of
the alluring traits of the YM theory, such as renormalizability, asymptotic freedom, non-trivial topological configurations, etc.

In 1920, Rudolf Bach proposed an action based on the square of the Weyl tensor $C_{\mu\nu\alpha\beta}$  where the Weyl tensor itself is an invariant under local re-scalings of the metric~\cite{Bach1}. In 1977 Kaku, Nieuwenhuizen, and Townsend~\cite{Kaku} showed that Bach's action is the action of the gauge theory of the conformal group  provided that  conformal boosts are gauged by means of a non-propagating gauge field. In other words, Bach's theory of gravity may be regarded as the gauge theory of the conformal group. In fact, paper~\cite{Kaku} was apparently the first paper that explicitly referred to Bach's action as Weyl gravity (WG) --- a terminology that we utilize throughout this paper.

As anticipated, quantum Weyl gravity (QWG) has proved to have a number of desirable features that are shared together with quantum YM theory. In particular, as in the YM theory the coupling constant $\alpha$ is dimensionless, and this makes the theory perturbatively renormalizable. In accordance with analogy with YM instantons, QWG has also gravitational instantons that encode information about a non-perturbative vacuum structure of QWG~\cite{Strominger,Hartnoll}. Particularly intriguing is a parallel between QWG and the $SU(3)$ YM gauge theory of the strong force (QCD). Some speculation about an analogy between quadratic gravity (including QWG) and QCD (which extends an old analogy between general relativity and the chiral Lagrangian of QCD) has occurred already before~\cite{Hill,Maggiore} but QWG is very particular in this respect.
Similarly to QCD, the gravitational interaction in QWG exhibits an antiscreening behavior at high energies on the account of the negative $\beta$-function. There are also strong indications that the ensuing UV fixed point of QWG is Gaussian~\cite{JLK,Percacci1} (as in QCD) and that the dynamical breakdown of the scale (Weyl) symmetry in QWG~\cite{JLK} in IR might be compared  to the confinement-deconfinement phase transition in QCD where it is anticipated that a mass gap (effective gluon mass) develops in the confining phase~\cite{Cornwall}.
In addition, certain aspects of the QCD functional integral and its measure are also shared by QWG~\cite{Holdom}.



Nevertheless, QWG is typically not considered as a viable candidate for a fundamental theory of quantum gravity for at least two reasons: i) its perturbative spectrum contains ghosts (namely a massless spin-$2$ dipole ghost) and, ii) its $\beta$-function at the one-loop level is non-zero, which implies a non-vanishing conformal (trace) anomaly. Since in QWG,
one has conformal symmetry in a gauged (local) version, the appearance of the anomaly is typically considered disastrous for the construction
of the quantum theory, because it signals the absence of the symmetry defining the theory, on the quantum level~\cite{shapiro2}. This should be contrasted with quite benign conformal anomaly that arises in YM theories.

The problem i) is not specific only to QWG but it is shared by all higher derivative gravity (HDG) theories. In these theories
the unitarity is in danger because there are perturbative states with negative kinetic energy or negative mass square parameter --- so-called ghosts or tachyonic states, respectively~\cite{BD}.
For QWG the problem can be most directly seen from the propagator
of the metric-field fluctuations around a flat background. The tensor structure
projects out spin-$2$ and spin-$1$ degrees of freedom. The
propagator in the spin-$2$ sector exhibits one massless
pole for normal graviton and another massless pole
with negative residue. The tachyonic states  can be easily eliminated from the spectrum of the theory by properly shifting the  vacuum state.
There are two possible interpretations of
the massless ghost pole, which depend on the prescription of the $i\varepsilon$ term. One can view it either as a state of negative norm or
a state of negative energy. In both cases, the optical theorem implies that the $S$-matrix based on the ensuing perturbation theory is inevitably non-unitary. Obviously, both ghosts and tachyons in HDG theories are undesirable and
various approaches have been invoked to remove them (or their effects) from the observable predictions of the theory: Lee--Wick prescription~\cite{Lee}, fakeons~\cite{Anselmi}, perturbative expansion around true vacuum state~\cite{JLK}, non-perturbative numerical methods~\cite{Tkach,Smilga,Tomboulis,Kaku2}, ghost~instabilities~\cite{Shapiro,Donoghue,Asorey}, non-Hermitian $PT$-symmetric quantum gravity~\cite{Bender}, etc. One might even entertain the idea that unitarity in quantum gravity is not a fundamental
concept~\cite{Hartle,Politzer,Lloyd}. So far, none of the proposed solutions has solved the problem conclusively.
Moreover, for QWG one could question whether perturbative non-unitarity is really an issue for an asymptotically free theory whose IR degrees of freedom are probably different from the ones in the UV.


Instead of debating various attitudes that can be taken toward the ghost/unitarity issue, our aim here is more modest.
We  will explore,  via functional renormalization group (FRG)~\cite{Reuter,Wetterich,codello} the  low-energy phenomenology of the QWG and see whether it can provide a realistic cosmology and what role (if any) is played by ghost fields.  We  start  from  the  UV  fixed point (FP) where the potential quantum gravity has an exact scale-invariance, so that only causal structure of events is relevant. This is a minimalistic assumption about quantum gravity in UV.  In order not to invoke any unjustified structure, we consider only metric-field based gravity without any matter field.  The UV FP in question might be, for instance, one of the critical points in a series of putative phase transitions that the Universe underwent in the very early (pre-inflationary) period of its evolution. Out of many scale-invariant HDG candidate theories, we choose to work with the simplest one, namely the theory that has only one single coupling constant.  The latter corresponds to the QWG theory.  Precisely at the UV FP,  the  QWG  has  exact  local  scale-invariance. A consistency of the entire scheme implies that  the UV  fixed point must be Gaussian~\cite{JLK}.
Existence of the Gaussian UV FP for QWG was also conjectured in earlier works~\cite{Percacci1,Percacci_R}.
We start with this premise  and  let  the  theory  flow  toward  IR energy scales.  In the close vicinity of the UV FP, the Weyl symmetry in the renormalized action is still preserved as only the Gauss--Bonnet ({\rm GB}) term is generated. Corrections explicitly violating Weyl symmetry,  such as  the $R^2$ term,  are generated  only  at  the second (or higher)  loop order.

In the vicinity of the UV FP, we choose a truncation ansatz for the effective action that will be further used to set up the FRG flow equation. Our truncation prescription is directly dictated by the one-loop effective action.  We further enhance this  by incorporating into the FRG equation  two non-perturbative effects, namely threshold  phenomena  and  the  effect  of  graviton anomalous dimension.
One may also wonder about how a consistent quantum theory can emerge when the action is problematic at tree-level (ghost problem). Nevertheless, the RG flow analysis reveals that by the inclusion of the two non-perturbative effects the quantum theory yields a sensible
IR FP. Indeed,  by solving the RG flow  equation  algebraically  for  $\beta$-functions  $\beta_{C}$ and $\beta_{{\rm GB}}$  we  show  that  there  exists  a  non-Gaussian  IR fixed point where both $\beta$-functions simultaneously disappear.  Aforementioned IR FP represents a critical point after which the (global) scale-invariance is broken. This is  reflected through the presence of  a composite  order-parameter  field  of  the Hubbard--Stratonovich  type, which  in the broken phase acquires a  non-trivial vacuum expectation value.

We map the broken-phase effective action on a two-field hybrid inflationary model that in its low-energy phase approaches the Starobinsky $f(R)$ model with a non-trivial gravi-cosmological constant. Requirement that Einstein's $R$ term in the low energy actions must have a coupling constant $1/2\kappa^2$ ties up the values of Starobinsky's inflation
parameter $\xi$ and the gravi-cosmological constant $\Lambda$. This fixes the symmetry-breakdown scale for QWG to be at about the GUT inflationary scale. Moreover, the existence of a regime where gravity is approximately scale invariant (fixed-point regime and departure of the RG flow from it) provides a simple and natural interpretation for the nearly-scale-invariance of the power spectrum of temperature fluctuations in the Cosmic Microwave Background radiation.

%

As for the conformal anomaly, typical imprints of it are new terms generated in the functional-integral action via higher loop corrections.
In particular, loop corrections will generate the Weyl-symmetry violating $R^2$ term in the action.
There are various possible ways how one can deal with conformal anomaly.
For instance, one might embed QWG in ${\cal N} =4$ conformal supergravity found by Fradkin and Tseytlin in 1985~\cite{FradkinT1,Duff} (which is known to be unique anomaly-free theory containing Weyl gravity in its bosonic sector) or in Witten--Berkovits twistor superstring theory~\cite{Berkovits:2004jj}. One should then consider QWG as the low-energy limit of such UV-finite models. In these cases, QWG is confined to the purely bosonic sector of spin-2 and spin-1 fluctuations where the conformal symmetry is only softly violated. Another option is to take seriously also theories with anomalous conformal symmetry similarly as done, for instance, in string theory with strings propagating in non-critical spacetime dimensions~\cite{Polchinski}. Here we will employ yet another scenario, in particular we will take advantage of the fact that the  symmetry-breaking $R^2$ term is presumably generated  only at the two-loop order (so that to one loop we do not observe the appearance of the ensuing anomalous conformal mode). On the other hand, our FRG approach will show that the IR FP
(for the two involved couplings) appears already at the enhanced one-loop level and hence the prospective observational consequences of the trace anomaly do not take over before the Weyl symmetry is dynamically broken. In the broken phase, the trace anomaly is not anymore a signal of inconsistency since the theory there is not scale-invariant to begin with and hence there is no reason why the ensuing energy momentum tensor should be traceless. 

The word {\em dark} in the title of this paper refers to two different things. First, it is related to conceptual and technical issues that plague QWG and that make it discouraging or {\em dark} in eyes of many practitioners. Second, appearance of the dynamical gravi-cosmological constant in the broken phase of QWG can be associated with ensuing {\em dark} energy.
It is purpose of this work to demonstrate that despite the aforementioned problems, QWG  may serve as a healthy theoretical setup for the UV-model  building  of  phenomenologically  viable  quantum theory of gravity with pertinent cosmological implications.


Our paper is organized as follows. In the next Section~\ref{1sab} we discuss some fundamentals of both classical and quantum Weyl gravity. In particular, we highlight a formal similarity of the WG with nonabelian Yang--Mills theories and  stress  some of  dissimilarities and potential  problems  met during quantization. Section~\ref{3sa} is dedicated to the construction of the FRG flow equation for the QWG in the one-loop
enhanced scheme, and with this tool we analyze in Section~\ref{4sc} the running of the $\beta$-functions associated  with the Weyl tensor square and Gauss--Bonnet terms.  In particular, we demonstrate that apart from the IR--stable fixed point that is reached at a zero-value of the running scale, the RG flow also exhibits a non-trivial bouncing behavior in the vicinity of the IR fixed point.
The issue of conformal anomaly of QWG is discussed in Section~\ref{tanomaly}. There we point out various technical and conceptual issued associated with the conformal anomaly in QWG and propose possible remedies.
In Section~\ref{5a} first employ  a  Hubbard--Stratonovich (HS) transformation,
which introduces  a  non-dynamical  {\em spurion} scalar  field  without  spoiling  the  particle  spectrum and (perturbative) renormalizability.  After the dynamical breakdown of the Weyl symmetry the HS field
acquires a non-trivial vacuum expectation value and gets radiatively generated gradient (kinetic) term.
If QWG has any physical relevance, then its effective action in the broken phase must contain an Einstein--Hilbert term. This is shown in
the second part of Section~\ref{5a}. We further demonstrate that in the broken phase the corresponding one-loop effective action consists (in the  Einstein  frame) of two  scalar fields --- scalaron and dynamical Hubbard--Stratonovich field, that  interact  via  derivative  coupling.
The resulting low-energy behavior in the broken phase can be identified with Starobinsky's $f(R)$-model (SM) with a gravi-cosmological constant (dark side of Weyl gravity)  that has a negative sign in comparison to the usual matter-induced cosmological constant. After this we
discuss  anomaly matching conditions between symmetric and broken phase of QWG.  A brief summary of results and  related  discussions  are  provided  in  Section~\ref{concl}.

\section{Some fundamentals of quantum Weyl gravity \label{1sab}}
\subsection{Classical Weyl gravity}
The WG is a pure metric theory that is invariant not only under the action of
the diffeomorphism group, but also under Weyl rescaling of the metric tensor  by the
local smooth functions $\Omega(x)$: $g_{\mu\nu}(x)\rightarrow \Omega^2(x)g_{\mu\nu}(x)$.
%
%
The simplest WG action functional in 4 spacetime dimensions that is both diffeomorphism and
Weyl-invariant has the form~\cite{Weyl1,Bach1,Mannheim:2017}
\begin{equation}
S \ = \ -\frac{1}{4\alpha^{\tiny{2}}}\int d^4x \ \! \sqrt{|g|} \ \! C_{\mu\nu\rho\sigma}
C^{\mu\nu\rho\sigma}\, ,
\label{PA1}
\end{equation}
where  $C_{\mu\nu\rho\sigma}$ is the
{\it Weyl tensor\/} which can be written as
\begin{eqnarray}
C_{\mu\nu\rho\sigma}  &=&
R_{\mu\nu\rho\sigma} \ - \ \left(g_{\mu[\rho}R_{\sigma]\nu} -
g_{\nu[\rho}R_{\sigma]\mu} \right)
 \ + \ \mbox{$\frac{1}{3}$}R \ \! g_{\mu[\rho}g_{\sigma]\nu}\, ,
\label{P2a}
\end{eqnarray}
with $R_{\mu\nu\rho\sigma}$ being the {Riemann curvature tensor},
$R_{\mu\rho}=R_{\mu\nu\rho}{}^\nu$  the  {Ricci tensor}, and $R= R_\mu{}^\mu$
the {scalar curvature}. Throughout the text we employ the time-like metric signature $(+,-,-,-)$  whenever  pseudo-Riemannian (Lorentzian) manifolds are considered.
The dimensionless coupling constant $\alpha$ is conventionally chosen
so as to mimic the YM action. On the other hand, in order to make connection with the usual RG practice,
it will be more convenient to consider the inverse of the coupling $\alpha^2$. 
In the following we denote the coupling in Eq.~(\ref{PA1}) as $\omega_{_C}$,
by employing the prescription $\omega_{_C} \equiv 1/(4\alpha^2)$.

Henceforth, we employ the following  notation: for the square of
the Riemann tensor contracted naturally, that is $R_{\mu\nu\rho\sigma}R^{\mu\nu\rho\sigma}$ we use the symbol
$R^2_{\mu\nu\rho\sigma}$, the square of the Ricci tensor $R_{\mu\nu}R^{\mu\nu}$ we denote by simply $R_{\mu\nu}^2$, the square of the Ricci
curvature scalar is always $R^2$, while for the Weyl tensor square $C_{\mu\nu\rho\sigma}C^{\mu\nu\rho\sigma}$
we employ a shorthand and schematic notation $C^2$. When the latter is treated as a local invariant (not under volume integral) in $d=4$ dimensions one finds the following expansion of the $C^2$ invariant into standard invariants quadratic in curvature
\begin{equation}
C^2 \ = \ R_{\mu\nu\rho\sigma}^2 \ - \ 2R_{\mu\nu}^2 \ + \ \mbox{$\frac{1}{3}$}R^2\,.
\label{weylsquare}
\end{equation}
We will also need another important combination of the quadratic curvature invariants, namely Gauss--Bonnet ({\rm GB}) term
\begin{equation}
\mathcal{G }\ = \ R_{\mu\nu\rho\sigma}^2 \ - \ 4R_{\mu\nu}^2 \ + \ R^2\, ,
\label{GBdef}
\end{equation}
which in $d=4$ is the integrand of the Euler--Poincar\'{e} invariant~\cite{Nakahara}
\begin{eqnarray}
\chi \ = \ \frac{1}{32\pi^2}\int d^4x \ \! \sqrt{|g|} \ \! \mathcal{G }\, .
\end{eqnarray}

In order to underline the similarity  with nonabelian YM theories  we use
the Riemann curvature tensor so that it is related to the usual
general relativistic one by
\begin{eqnarray}
R_{\mu\nu\lambda}{}^\kappa  \ = \ R^\kappa{}_{\mu\nu\lambda}|_{\rm genrel}\, .
\end{eqnarray}
Note that due to skew and interchange symmetries of the Rieman tensor we have
 \begin{eqnarray}
 R_{\lambda\nu} \ = \ R_{\lambda\nu}|_{\rm genrel}\; \; \; \Rightarrow \;\;\; R  \ = \ R|_{\rm genrel}\, .
 \end{eqnarray}
 In addition, also
 \begin{eqnarray}
 C_{\lambda\mu\nu\kappa}
 C^{\lambda\mu\nu\kappa} \ = \  C_{\lambda\mu\nu\kappa}
 C^{\lambda\mu\nu\kappa}|_{\rm genrel}\, ,
 \end{eqnarray}
 trivially holds. In particular, in our convention we consider both Christoffel symbols and curvature tensors as
 $4\times4$ matrices; $R_{\mu\nu\lambda}{}^\kappa  \equiv  \{R_{\mu\nu}\}_{\lambda}{}^\kappa$ and
 $\Gamma_\nu{}_\lambda{}^\kappa  \equiv   \{ \Gamma_\nu\}_\lambda{}^\kappa$. With this one can write
 \begin{equation}
 R_{\mu\nu\lambda}{}^\kappa   \ = \ \{\partial_\mu \Gamma_\nu-\partial_\nu \Gamma_\mu-[ \Gamma_\mu,
  \Gamma_\nu]\}_\lambda{}^\kappa\, ,
 \label{P2}
 \end{equation}
which is clearly analogous to the relation
\begin{eqnarray}
F_{\mu \nu}^a \ = \ \{\partial_\mu A_\nu-\partial_\nu A_\mu - i\mbox{g} [ A_\mu,
  A_\nu]\}^a\, ,
\end{eqnarray}
for the field strength in  nonabelian YM theories with $A_{\mu}$ representing the gauge field.
A natural geometrical language for the description of this analogy is provided by the fiber-bundle theory.
While $\Gamma_\nu{}_\lambda{}^\kappa$ represents connection in the frame bundle, $A_\nu^a$  plays the same role in
the ensuing principal bundle. Similarly, $R_{\mu\nu\lambda}{}^\kappa$ quantifies  anholonomy  in a parallel transport around a closed infinitesimal loop in the frame bundle, whereas $F_{\mu \nu}^a$ quantifies the anholonomy in
the principal bundle. Although the fibre-bundle theory provides the most adequate language for the parallelism between Riemannian-geometry-based gravity and nonabelian YM theories
we will not pursue this point any further here.

%

With the help of the Chern--Gauss--Bonnet  theorem one
can cast the Weyl action $S$ into equivalent form (modulo topological term)
\begin{equation}
S \ = \
-\frac1{2\alpha^2}\int d^4x\ \! \sqrt{|g|} \ \!\left(R_{\mu\nu}^2 -
\mbox{$\frac{1}{3}$} R^2\right)\, .
\label{PA2}
\end{equation}
It should be stressed that although the omitted topological term is clearly not important on the classical
level, it is relevant on the quantum level where summation over distinct topologies
should be considered (see following subsection). But even when one stays on topologies with a fixed Euler--Poincar\'{e} invariant, the renormalization will
inevitably generate (already at one loop) the {\rm GB} term with a running coupling constant (see, e.g., Section~\ref{3sa}).

In passing we note also that both (\ref{PA1}) and (\ref{PA2}) are Weyl-invariant only in $d=4$ dimensions.
In fact, under the  Weyl transformation $g_{\mu \nu} \to \Omega^{2} {g}_{\mu \nu}$ and the densitized $C^2$
transforms as
\begin{eqnarray}
\sqrt{|g|} \ \! C^2 \ \to \ \Omega^{d-4} \sqrt{|g|}   C^2\, ,
\end{eqnarray}
in general dimension $d$ of spacetime, 
while $\sqrt{|{g}|} \ \!\mathcal{G}$  supplies topological invariant only in $d=4$.
This is particularly important to bear in mind during the quantization where (similarly as in the Yang--Mills theories) one should  choose such a regularization method that preserves the local gauge symmetry of the underlying Lagrangian and thus does not introduce any unwanted symmetry breaking terms. For this reason
one should preferentially rely on fixed-dimension renormalization and avoid, e.g., dimensional regularization. This is the strategy we will pursue also in this paper.


Variation of $S$ with respect to the metric
yields the field equation of motion (EOM) known as the Bach vacuum equation:
\begin{equation}
[2C^{\mu\lambda\nu\kappa}_{\phantom{\lambda\mu\nu\kappa};\lambda;\kappa}
-C^{\mu\lambda\nu\kappa}R_{\lambda\kappa}]
\ \equiv \ B^{\mu\nu}
\ = \  0\, ,
\label{Z42A}
\end{equation}
where $B^{\mu\nu}$  is the {\it Bach tensor\/} (trace-free tensor of rank $2$) and ``${;{\alpha}}$'' denotes the usual
covariant derivative (with Levi--Civita connection). We remind that the form (\ref{Z42A}) of the EOM is specific only to $d = 4$. Moreover, apart from being traceless, the Bach tensor is also divergence-free ($B^{\mu\nu};\mu=0$), which ia a consequence of diffeomorphism symmetry.
Because Bach tensor results from variational derivative of the action $S$ with respect to symmetric metric tensor $g_{\mu\nu}$, it must also be symmetric ($B^{\nu\mu}=B^{\mu\nu}$).
When for a given background $B^{\mu\nu}=0$ (i.e., given backgound is a classical vacuum solution of Weyl gravity), then we say that it is Bach--flat. More general discussion of classical singularity-free solutions in WG can be found in \cite{confreview,spcompl0,finconfqg}.
%
%

\subsection{Quantization\label{quantization}}
%
One can formally quantize WG by emulating the strategy known from quantum field theory, i.e.
by introducing a functional integral ($\hbar =  c = 1$)
\begin{equation}
Z \ = \ \sum_i \int_{\mathcal{M}_i} {\cal D}g_{\mu\nu} \ \! e^{iS}\, .
\label{Z42Ab}
\end{equation}
Here  ${\cal D}g_{\mu\nu}$ denotes the functional-integral measure
whose proper treatment involves  the Faddeev--Popov gauge
fixing of the gauge symmetry Diff$\times$Weyl$({\mathcal{M}}_i)$
plus ensuing Faddeev--Popov (FP) determinant~\cite{Tseytlin,Hamber}.
As for local factors $[-\det g_{\mu \nu}(x)]^{w}$ in the measure, we
choose to work with DeWitt convention~\cite{DeWitt1,DeWitt2}: $w = (d-4)(d+1)/8$. Clearly,
when the fixed-dimension renormalization scheme in $d=4$ is employed the
local factor does not contribute.
Strictly speaking, for a full-fledged quantization program one should consider generator of correlation functions $Z[J_{\mu\nu}]$,
i.e. functional of a source field that is coupled to a dynamical metric field. Such object would contain information on all
correlations functions but, unfortunately, it is far beyond our present computational capability. Fortunately, for our purposes, i.e.,
for the computation of enhanced one-loop effective action the use of quantum partition function (\ref{Z42Ab}) will suffice.

The sum in (\ref{Z42Ab}) is a sum over four-topologies, that is, the sum over topologically distinct
manifolds ${\mathcal{M}}_i$ (analogue to the sum over genera
in string theory or sum over homotopically inequivalent vacua in the Yang--Mills theory) which
can potentially contain topological phase factors, e.g., the Euler--Poincar\'{e} characteristic
of ${\mathcal{M}}_i$, cf. Refs.~\cite{Carlip}. It should be stressed that
the sum over four-topologies is a problematic concept since four-manifolds are generally un-classifiable
--- i.e., there is no algorithm that can determine whether two arbitrary four-manifolds are homeomorphic.
On the other hand, simply connected compact topological four-manifolds are classifiable in terms of {\it Casson handles}~\cite{Freedman},
which can be applied in functional integrals in Euclidean gravity.
%
%
For simplicity, we will further assume that all global topological effects can be ignored,
so, in particular, we assume  that our space ${\mathcal{M}}_i$
is compact and its tangent bundle is topologically trivial.

To avoid issues related to renormalization of non-physical sectors (i.e., Faddeev--Popov ghosts and longitudinal components of the metric field) it will be
convenient in our forthcoming reasonings to employ the York decomposition of the metric fluctuations
$h_{\mu\nu}$ defined as
\begin{eqnarray}
g_{\mu\nu}\ = \ g^{(0)}_{\mu\nu} \ + \ h_{\mu\nu}\,,
\end{eqnarray}
where we have denoted the background metric as $g^{(0)}_{\mu\nu}$. The York decomposition is
then implemented in two steps~\cite{Percacci_R}. In the first step
we rewrite the metric fluctuations as
%
\begin{eqnarray}
h_{\mu\nu}\ = \ \bar{h}_{\mu\nu} \ + \ \mbox{$\frac{1}{4}$}g_{\mu\nu}h\, ,
\end{eqnarray}
where $h$ is a trace part of $h_{\mu\nu}$ and  $\bar{h}_{\mu\nu}$ is the corresponding  traceless part.
More specifically,
\begin{eqnarray}
g^{(0)\mu\nu}\bar{h}_{\mu\nu} \ = \  \bar{h}_{\mu}{}^{\mu}\   = \  0\, , \quad \quad
h  \ = \  g^{(0)\mu\nu}h_{\mu\nu} \  = \  h_{\mu}{}^{\mu}\, .
\end{eqnarray}
We will always tacitly assume that the Lorentz indices are raised or lowered
via the background metric, i.e. via $g^{(0)\mu\nu}$ or $g_{(0)\mu\nu}$, respectively.
Also all covariant derivatives
$\nabla_\mu$ below will be understood as taken with respect to the background metric. In the following the operator $\Box$ will denote the so-called Bochner Laplacian operator~\cite{Percacci_R}, i.e., the covariant operator defined as $\Box\equiv\nabla^\mu\nabla_\mu$.

In the second step, we decompose the traceless part into the transverse, traceless
tensor $\bar{h}_{\mu\nu}^\perp$  and to parts carrying the longitudinal (i.e., unphysical) degrees of freedom, namely
\begin{eqnarray}
\bar{h}_{\mu\nu} \ = \ \bar{h}_{\mu\nu}^{\perp}\  + \ \nabla_{\mu}\eta_{\nu}^{\perp}\  + \ \nabla_{\nu}\eta_{\mu}^{\perp}\ + \  \nabla_{\mu}\nabla_{\nu}\sigma \  - \ \mbox{$\frac{1}{4}$}g_{\mu\nu}\square\sigma\, .
\label{Ydecomp}
\end{eqnarray}
These mixed-longitudinal (and traceless) parts are written in terms of an arbitrary transverse vector
field $\eta_\mu^\perp$ and a scalar (trace) degree of freedom $\sigma$. The last fields must
satisfy the usual conditions of transversality and  tracelessness, i.e.
\begin{eqnarray}
\nabla^{\mu}\bar{h}_{\mu\nu}^{\perp}\ = \ 0,\quad\nabla^{\mu}\eta_{\mu}^{\perp} \ = \ 0,
\quad\bar{h}_{\,\,\mu}^{\perp\mu} \ = \ 0\, .
\end{eqnarray}
The true propagating degrees of field in QWG are associated with the transverse and traceless field $\bar{h}_{\mu\nu}^{\perp}\equiv h^{TT}_{\mu\nu}$. Indeed, from the second variation of the
Weyl action  expanded around a generic background it can be seen that $\bar{h}_{\mu\nu}^{\perp}$ is the only field component that propagates on quantum level.
The vector field $\eta_\mu^\perp$ and two scalar fields $h$ and $\sigma$ completely drop out from the expansion due to diffeomorphism and conformal invariance respectively~\cite{Irakleidou}. Some explicit examples will be given in Section~\ref{4sc}.


\section{Exact RG flow for quantum Weyl gravity \label{3sa}}
%

For convenience sake, our subsequent reasonings will be done in
an $d=4$ Euclidean space dimensions as this is a typical framework in which the FRG treatment is done.
By performing Wick rotation from Minkowski space to Euclidean space the question of the resulting metric signature arises. When one does, in a standard way, only the change of the time coordinate $t\to -it_E$, (where $t_E$ is the name of the first coordinate in the Euclidean characterization of space) the resulting signature of the metric of space is completely negative, that is $(-,-,-,-)$. It seems natural to define the corresponding GR-covariant d'Alembert operator as $\square_E=-\nabla^\mu\nabla_\mu$, where the generalization to curved Euclidean space is done by using Bochner Laplacian. However, in all formulas that follow, we find more convenient to use the following definition in the Euclidean signature $\square=\nabla^\mu\nabla_\mu$. We also remark that this last operator $\square$, if analyzed on the flat space background has negative semi-definite spectrum. We will also use a definition of the covariant Euclidean box operator (covariant Laplacian) $\Delta=\square=-\square_E$ and this last operator in the Euclidean flat space case has a spectrum which is characterized by $-k^2$, the $d=4$ Euclidean negative square of a $4$-momentum vector $k_\mu$. Accordingly, the signature of the metric in Euclidean space  will be taken to be $(+,+,+,+)$.

The aim of this section is to explore the IR behavior of the QWG by starting from the presumed UV FP where the QWG is exact. Existence of such a  UV FP was self-consistently checked in Ref.~\cite{JLK}. To this end we will solve the FRG flow equation~\cite{Reuter,Wetterich}
for the effective average action $\Gamma_k$, which reads
\begin{eqnarray}
\partial_t \Gamma_k \ = \ \frac{1}{2} \mbox{Tr} \left[\partial_t R_k (\Gamma^{(2)} + R_k)^{-1} \right].
\label{FRG_1a}
\end{eqnarray}
 The IR-cutoff $R_k$ suppresses the contribution of modes with small eigenvalues of the covariant Laplacian
$-\Delta \ll k^2$,  while the factor $\partial_t R_k$ removes contributions from large eigenvalues of $-\Delta \gg k^2$.  In this way the loop integrals are both IR- and UV-finite~\cite{Wetterich2}. The second variational derivative of the effective action --- $\Gamma^{(2)}$, depends on the background metric $g_{\mu \nu}^{(0)}$, which is the argument of the running effective action
$\Gamma_k$, while $k$ is the running energy (momentum) scale or the momentum of a mode in the Fourier space. We also employ the notational convention $\partial_t = k \partial_k$.

Ideally, Eq.~(\ref{FRG_1a}) would require calculation of the full resummed and RG-invariant effective action.
It is, however, difficult  to proceed analytically in this way  so we content ourselves here with
the conventional procedure, according to which one should employ some well motivated ansatz for the effective action.
In particular, in order to evaluate the RHS of Eq.~(\ref{FRG_1a}) we employ the
(Euclidean) effective action in the enhanced one-loop scheme. By the enhanced one-loop scheme we mean one-loop effective action in which also effects of the anomalous dimension
and threshold phenomena are included. Corresponding truncation will thus inevitably go beyond the usual polynomial ansatz.
On the other hand, for the LHS of~(\ref{FRG_1a}) we project the flow on the subspace of the three invariants containing precisely four derivatives of the metric (Eqs. (\ref{weylsquare}), (\ref{GBdef}) and $R^2$ invariant).
The reason why we consider effective action on the RHS being different from the effective action on the LHS, is dictated by technical convenience. Namely, the RHS acts as source for the RG flow, while the LHS
contains the desired structure of the effective action that is appropriate for the extraction of the $\beta$-functions.

Let us now briefly describe the basic steps that are used in solving the FRG flow equation. Further technical details as well as necessary derivations can be found in our recent paper~\cite{JLK} and in its supplemental material~\cite{supplement}. On the other hand, our final results will be discussed in more detail in the following section.

First step is the construction of the one-loop partition function for QWG. It is important here to choose convenient class of backgrounds. We selected maximally symmetric spaces (MSS) and Ricci-flat backgrounds, knowing that both of them are also Bach-flat, so they are classical exact solutions to the Bach equation  (i.e., vacuum equations in WG). The partition function we construct for physical degrees of freedom (only transverse and traceless gravitons) and for this we use York decomposition outlined above. We take care of the change of variables under the functional integral, addition of the Faddeev--Popov determinant, inclusion of the differential Jacobian, and also on exclusion of the contribution of zero modes, which are unwanted. We also check that all these one-loop partition functions provide 6 propagating degrees of freedom in QWG.

In the second step, we analyze the general $\beta$-functional of the theory considered on our general backgrounds (MSS and Ricci-flat). We ask and address the question about information which can be extracted on these backgrounds about the $\beta$-functions of involved couplings. Due to various relations between curvature tensors and between square of tensors, we find that we are able to extract only two combinations of couplings: either $\beta_R+\frac{1}{6}\beta_{{\rm GB}}$ on MSS, or $\beta_C+\beta_{{\rm GB}}$ on Ricci-flat background. It can be also checked that the usage of general Einstein spaces as on-shell backgrounds does not improve on this situation. However, in a general theory to the quadratic order in curvatures we may set up an ansatz for the effective average action $\Gamma$ built out of three couplings $\omega_C$, $\omega_{{\rm GB}}$, and $\omega_R$ corresponding to three quadratic invariants $C^2$ term, $\mathcal{G}={\rm {\rm GB}}$ term, and $R^2$ term. We address this potential inconsistency of the RG system (three couplings and only two extractable $\beta$-functions) in the final step of our method of solving the FRG.

The most important part consists of building and solving the FRG flow equation. In Ref.~\cite{JLK} we found a novel form of the FRG flow equation based on the expressions for factors appearing in the quantum partition function of the theory. These factors mimic the simple scalar two-derivative kinetic sectors of the theory, however they may appear both in the numerator and in the denominator of the partition function and with various mass square parameters (which can also be negative). We take into account quantum wave-function renormalization of quantum fields and add to the flow equation the anomalous dimension $\eta$ for all fields participating in the quantum dynamics at the one-loop level. To do the IR-suppression of modes in each factor we should add a suitable cutoff kernel function $R_k(z)$. When all these operations are done the final form of the FRG flow equation, reads
\begin{equation}
\partial_{t}\Gamma_{L,k} \ = \ \frac{1}{2}\sum_i \sum_j \ \pm \ {\rm Tr}_{\phi_i}\left(\frac{\left(\partial_{t}R_{k}-\eta R_{k}\right)\hat{1\hspace{-1mm}{\rm I}}}{\hat{\square}+R_{k}\hat{1\hspace{-1mm}{\rm I}}
-Y_{i,j}}\right),
\label{FRG_partition}
\end{equation}
where $Y_{i,j}$ are  mass square parameters of the modes $\phi_i$ and the functional trace is done over space of the same modes, which may also contain some traces over internal indices of the fields. The frontal plus/minus sign in (\ref{FRG_partition}) originates from the initial position of the factor in the partition function (whether it was in the denominator, or in the numerator of the partition function, respectively).

The final step consists of choosing the truncation ansatz for the action, whose FRG flow we try to determine. To write the RHS of the flow equation we explicitly reflect the fact that our theory is  QWG theory, where we could add a topological {\rm GB} term, therefore not influencing at all the perturbative one-loop partition function. On the LHS we could analyze the flow of the general action quadratic in curvatures with the structure of the Lagrangian $\omega_C C^2+\omega_{{\rm GB}} \mathcal{G}+\omega_R R^2$. However, in perturbative QWG at the one-loop level there is a very interesting and special simplification since at this quantum level the $R^2$ term is not needed, and we know that perturbatively its one-loop $\beta$-function $\beta_R$ vanishes. We utilize this hierarchy of $\beta$-functions in QWG and employ consistently the truncation ansatz without the problematic $R^2$ term. Therefore we have two combinations of $\beta$-functions ($\frac{1}{6}\beta_{{\rm GB}}$ and $\beta_C+\beta_{{\rm GB}}$) and two couplings $\omega_C$ and $\omega_{{\rm GB}}$, so the system is consistent and possible to be solved algebraically for the $\beta$-functions involved.

This solving we do in the last technical step, where we evaluate all the functional traces (both in internal space and spacetime volume integrals). We perform the traces using the heat kernel  method and Barvinsky--Vilkovisky trace technology. In order to have an analytic control over all formulas, we decide to choose a particular form of the cutoff kernel function $R_{k} \ \equiv \  R_{k}(z)\ = \ \left(k^{2}-z\right)\theta\!\left(k^{2}-z\right)$ due to Litim~\cite{Litim2}. The main point of using the Wetterich equation (\ref{FRG_1a}) is to take into account massive modes which slow down the RG flow in the IR regime. We achieve this by adding cutoff kernels $R_k$ in mass-dependent renormalization scheme of Wilsonian character. This lets us to obtain one-loop RG-improved expressions for the two $\beta$-functions of the theory with all quantum effects due to anomalous dimension $\eta$ and IR threshold phenomena included.


\section{Analysis of $\beta$-functions and RG fixed points \label{4sc}}

We now discuss the system of $\beta$-functions of the theory considered at the one-loop level improved by the usage of FRG methods. Based on the computation presented in Ref.~\cite{JLK} the explicit form of  $\beta$-functions reads

\begin{eqnarray}
\beta_{\rm {\rm GB}}\! &=&\! \frac{1}{2}(2-\eta)\left[-\frac{21}{40}\left(1-\frac{\frac{2}{3}\Lambda}{k^{2}}\right)^{-1}
\ + \ \frac{9}{40}\left(1-\frac{\frac{4}{3}\Lambda}{k^{2}}\right)^{-1} \ - \ \frac{179}{45}\left(1+\frac{\Lambda}{k^{2}}\right)^{-1}\ -\ \frac{59}{90}\left(1+\frac{\frac{1}{3}\Lambda}{k^{2}}\right)^{-1}\right.\nonumber \\[1mm]
 &&+ \ \left. \ \frac{479}{360}\left(1+\frac{2\Lambda}{k^{2}}\right)^{-1}\ - \ \frac{269}{360}\left(1+\frac{\frac{4}{3}\Lambda}{k^{2}}\right)^{-1}\right], \label{betasys1}
\end{eqnarray}
and
\begin{eqnarray}
\mbox{\hspace{-0mm}}\beta_{C} \ + \ \beta_{{\rm GB}} \ = \frac{2-\eta}{2}\frac{137}{60}\, ,
\label{betasys2}
\end{eqnarray}
with the anomalous dimension of the graviton field given by
\begin{equation}
\eta \ = \ -\frac{1}{\omega_{C}}\ \!\beta_{C}\, ,
\label{73abc}
\end{equation}
to the one-loop level of accuracy. By $\omega_{C}=\omega_{C}(k)$ we here denote a running coupling parameter
in front of the $C^{2}$ term in the action (\ref{PA1}) --- the so-called Weyl coupling.

The above two $\beta$-functions, $\beta_{C}$ and $\beta_{{\rm GB}}$, follow from the FRG  with the truncation ansatz motivated by the one-loop level and with both the threshold phenomena and non-trivial anomalous dimension of the quantum graviton included. We call them thus as being one-loop RG-improved. The effects of threshold phenomena are present explicitly only in the expression (\ref{betasys1}) for $\beta_{{\rm GB}}$. However, due to the combination in (\ref{betasys2}), the solution for $\beta_C$ will also inherit these threshold factors. Finally, we observe that  the anomalous dimension $\eta$ enters only multiplicatively in the system of $\beta$-functions (\ref{betasys1})-(\ref{betasys2}). This has some simplifying consequences for a search for FP's of the coupled system, both in the UV  as well as in the IR regimes.

Let us now discuss the reasons for the presence of threshold phenomena in our system. As it could be seen from the expressions for the one-loop partition functions of the system \cite{JLK} on MSS,
\begin{eqnarray}
\mbox{\hspace{-5mm}}Z_{{\rm 1-loop}}^{2} &=&\frac{\det^{2}_1\left(\hat\square+\Lambda\hat{1\hspace{-1mm}{\rm I}}\right)\det_{1}\left(\hat\square+\frac{1}{3}
\Lambda\hat{1\hspace{-1mm}{\rm I}}\right)\det_{0}\left(\hat\square+\frac{4}{3}\Lambda\hat{1\hspace{-1mm}{\rm I}}\right)}{\det_{2T}\left(\hat\square-\frac{2}{3}
\Lambda\hat{1\hspace{-1mm}{\rm I}}\right)\det_{2T}\left(\hat\square-\frac{4}{3}\Lambda\hat{1\hspace{-1mm}{\rm I}}\right)\det_{0}\left(\hat\square+2\Lambda\hat{1\hspace{-1mm}{\rm I}}\right)}
\, ,
\label{17ab}
\end{eqnarray}
and on Ricci-flat background
\begin{eqnarray}
Z_{{\rm 1-loop}}^{2}=
\frac{{\rm det}^3_{1}\hat\square\,{\rm det}_{0}^2\hat\square}{{\rm det}^2_{2}\left(\hat\square-2\hat{C}\right)}\,, \label{Ricflatpartition}
\end{eqnarray}
the box-kinetic operator of all quantum modes is shifted only in the case of MSS background. This is the reason to produce IR thresholds. The shift by a matrix of a Weyl tensor $\hat C$ on the Ricci-flat background does not generate any threshold because of the tracelessness of the Weyl tensor. These shifts in the factors in the partition function (\ref{17ab}) on MSS backgrounds are analogous to massive modes in standard QFT. Their role is to effectively slow down the RG flow in the IR regime since there the quantum fields become heavy (with mass).

We can further simplify the system of equations (\ref{betasys1})-(\ref{betasys2}) for the two $\beta$-functions. In particular, we do not wish to solve explicitly the system (\ref{betasys1})-(\ref{betasys2}). We just concentrate on the corresponding FP's. This is a much simpler task as we can solve the system of $\beta$-functions algebraically. This gives
\begin{equation}
\beta_{C} \ = \ \frac{b-{\cal X}}{1+y({\cal X}-b)}, \;\;\;\; \beta_{{\rm GB}} \ = \ \frac{{\cal X}}{1+y({\cal X}-b)}\, ,
\label{71ac}
\end{equation}
where
\begin{eqnarray}
\mbox{\hspace{-5mm}}{\cal X} &=& - \ \frac{21}{40}\left(1-\frac{\frac{2}{3}\Lambda}{k^{2}}\right)^{-1} \ + \ \frac{9}{40}
\left(1-\frac{\frac{4}{3}\Lambda}{k^{2}}\right)^{-1}-  \frac{179}{45}
\left(1+\frac{\Lambda}{k^{2}}\right)^{-1} \ - \ \frac{59}{90}\left(1+\frac{\frac{1}{3}\Lambda}{k^{2}}\right)^{-1}\nonumber \\[1mm]
&+ & \frac{479}{360}\left(1+\frac{2\Lambda}{k^{2}}\right)^{-1} \ - \
\frac{269}{360}\left(1+\frac{\frac{4}{3}\Lambda}{k^{2}}\right)^{-1},
\label{Xeqn}
\end{eqnarray}
with $b  =  {137}/{60}$ and $y={1}/{\omega_{C}}$. The origin of the common denominator $1+y({\cal X}-b)$ is entirely due to the inclusion
of the graviton's anomalous dimension $\eta$. When $\eta$ is neglected, the latter is unity. This is the regime, in which the Weyl
coupling $\omega_C$ is big ($\omega_C\to\infty$, so $y\to0$). This corresponds to the perturbative regime of the theory (in terms of $\alpha_C$). When one decides to
neglect these common denominators, one gets  simplified expressions,
\begin{eqnarray}
\beta_{C} \ = \ b-{\cal X}, \;\;\;\; \beta_{{\rm GB}} \ = {\cal X}\, ,
\label{simplsys}
\end{eqnarray}
which are already sufficient to shed light on the issue of existence and character of FP's of the FRG flow. The equations in (\ref{simplsys}) still include the effects of threshold phenomena. When we neglect the threshold phenomena in our description,
then the system  of $\beta$-functions acquires the form of one-loop perturbative system as derived
in~\cite{Tseytlin} for QWG in dimensional regularization scheme. Actually, all these threshold phenomena are contained in the expression called ${\cal X}$ above. When one takes the limit $\Lambda/k^2$ to zero,
then all threshold factors are indeed removed, and the expression ${\cal X}$ reduces to  just a number $\beta_{{\rm GB}}^{\rm FT}={-87}/{20}$.
%

\subsection{Ultraviolet Asymptotic Freedom in all couplings}\label{UVAF}

 When $\kappa = k/\sqrt{|\Lambda|} \gg 1$, the threshold phenomena are completely irrelevant and can be neglected, cf. Eq.~(\ref{betasys1}).  Irrespectively of the initial values of the couplings $\omega_{C0}=\omega_C(t_0)$ and $\omega_{GB0}=\omega_{{\rm GB}}(t_0)$ (look also at an analysis in the next paragraph),
the leading RG running behavior
 in the UV regime (for $t\gg1$) is $\omega_C\sim t \beta_C^{\rm FT}$ and $\omega_{{\rm GB}}\sim t \beta_{{\rm GB}}^{\rm FT}$. This signifies that the absolute values of the couplings
must necessarily grow in the UV.
It might be argued that in the UV regime one can also
neglect the effects of the anomalous dimension $\eta$, cf.~Eq. (\ref{73abc}), since it is suppressed by big values of the $\omega_C$ coupling in the UV.
For the UV-running ($t\to+\infty$ equivalent to $k\gg k_0$), it suffices to use only the non-RG-improved one-loop perturbation results (\ref{simplsys})  from above. 

Fradkin and Tseytlin~\cite{Tseytlin} were the first to find that  the one-loop
$\beta$-functions for $\omega$-couplings are constants with values
\begin{eqnarray}
\beta_{C} \ = \ \frac{199}{30}\quad{\rm and}\quad\beta_{{{\rm GB}}}\ = \ -\frac{87}{20}\, .\label{betace}
\end{eqnarray}
This leads to an asymptotic freedom at the UV FP for all two couplings as
we shall prove below. Since $\beta_{C}>0$, it is natural to assume
that the initial condition of the flow is such that $\omega_{C}\left(t_{0}\right)>0$
and similarly since $\beta_{{{\rm GB}}}<0$, then $\omega_{{{\rm GB}}}\left(t_{0}\right)<0$.
If one chooses the opposite condition, then the RG flow tends to decrease
the absolute value of the $\omega$-coupling, the coupling crosses
zero, and finally it goes on the other side, where the initial conditions
are natural in a sense mentioned above. This is because one-loop RG
flow in the UV forces the $\beta$-functions to be constants, so the increments
of the couplings (positive for $\omega_{C}$ and negative for $\omega_{{\rm GB}}$)
are regular and linear in the UV regime. Hence, in the deep UV (close
to the UV FP) we can assume that $\omega_{C}>0$ and also that $\omega_{{\rm GB}}<0$.

All these arguments are self-consistent and lead to the conclusion that the UV fixed point of RG inevitably exists and realizes the asymptotic freedom (AF) scenario
(in much the same way as in non-Abelian gauge theories). Our perturbation analysis is carried out in terms of the coupling $g^2 \propto {1}/{\omega}$, so this and
the fact that in the UV regime, $\omega_C\to+\infty$, bolsters even more the
correctness of our perturbative one-loop results. Actually, near the UV Gaussian FP, it is the coupling $g$
(analogous to the YM coupling constant)
that goes to zero.

The asymptotic freedom (AF) characterization of the FP comes when the RG
flow is analyzed in terms of  $g$-like couplings. We define them, taking into
account above signs of $\omega_{C0}$ and $\omega_{GB0}$, in the following way:
\begin{eqnarray}
\omega_{C}\ = \ \frac{1}{g_{C}^{2}} \ = \ \frac{1}{4\alpha^2} \quad{\rm and}\quad\omega_{{\rm GB}} \ = \ -\frac{1}{g_{{{\rm GB}}}^{2}}\, ,
\end{eqnarray}
and with the inverse relations
\begin{eqnarray}
g_{C} \ = \ \frac{1}{\sqrt{\omega_{C}}}  \quad{\rm and}\quad g_{{{\rm GB}}}\ = \ \frac{1}{\sqrt{|\omega_{{\rm GB}}|}} \ = \ \frac{1}{\sqrt{-\omega_{{\rm GB}}}}\,.
\end{eqnarray}
In this way we are sure that both $g$-couplings are non-negative.
Now, one can easily derive by differentiation
\begin{eqnarray}
\beta_{C} \ = \ \partial_{t}\omega_{C}\ = \ -\frac{2}{g_{C}^{3}}\partial_{t}g_{C}\ = \ -\frac{2}{g_{C}^{3}}\beta_{g_{C}}\, ,
\end{eqnarray}
and
\begin{eqnarray}
\beta_{{\rm GB}} \ = \ \partial_{t}\omega_{{\rm GB}}=\frac{2}{g_{{{\rm GB}}}^{3}}\partial_{t}g_{{{\rm GB}}}=\frac{2}{g_{{\rm {\rm GB}}}^{3}}\beta_{g_{{{\rm GB}}}}\, ,
\end{eqnarray}
(mind the sign in the second equation). Then the $\beta$-functions
of these couplings are expressed as
\begin{eqnarray}
\beta_{g_{C}} \ = \ -\frac{1}{2}g_{C}^{3}\beta_{C}\quad{\rm and}\quad\beta_{g_{{\rm {\rm GB}}}} \ = \ \frac{1}{2}g_{{{\rm GB}}}^{3}\beta_{{\rm GB}}\,,
\end{eqnarray}
which to one-loop accuracy read
\begin{eqnarray}
\beta_{g_{C}} \ = \ -\frac{199}{60}g_{C}^{3}\quad{\rm and}\quad\beta_{g_{{{\rm GB}}}}\ = \ -\frac{87}{40}g_{{{\rm GB}}}^{3}\,.\label{betasys1loopg}
\end{eqnarray}
For vanishing values of the $g$-couplings we find that the above $\beta$-functions vanish too, so in this situation we have a trivial Gaussian FP of the RG flow.
We also see that for the positive values of the couplings, that is $g_{C}>0$
and $g_{{{\rm GB}}}>0$, we have that both $\beta$-functions $\beta_{g_{C}}$
and $\beta_{g_{{{\rm GB}}}}$ are negative which signifies that in the
UV we meet a  FP (asymptotically free theory) in basically
the same way like this happens in QCD.

\subsection{Scheme independence of IR fixed point}

In the paper \cite{JLK}, the detailed analysis of the situation with FP's in the IR was presented. We found both: a turning point of the RG flow at some finite energy scale $k$ and also a true IR FP at $k=0$, so in the deep IR regime. The genuine IR FP has a non-trivial (non-Gaussian) character and it lets to define the theory in the non-perturbative way, which is free from any IR-type of divergences and it therefore realizes Weinberg asymptotic safety, but in the infrared. The turning point (TP) of the RG flow is related to the non-analytic behavior of the $\beta$-functions in terms of running coupling parameters. This in turn corresponds to a very interesting cosmological bounce scenario, when analyzed from the AdS/CFT correspondence point of view. In this subsection, we discuss some universal features of the TP of RG and of its location, while in the next  subsection \ref{s43}, we prove the relation between the non-analytic behavior of $\beta$-functions and the bounce happening at finite $k$.

One can easily see that qualitative features of the RG flow presented
above do not depend on the details of the regularization and renormalization
procedures. For example, taking a closer look at the plot (cf. Fig.1. from \cite{JLK}) of the dependence
of the $\beta$-functions on the energy scale, one can convince oneself
that the asymptotics in the UV limit of the flow and an existence
of the location of the first zero of the RG flow counting from UV direction are universal.
Firstly, we would like to discuss the UV regime of the RG flow.
The asymptotics in the UV limit is described by one-loop $\beta$-functions
computed by Fradkin and Tseytlin for 4-dimensional conformal gravity
and our exact (or rather FRG-improved) $\beta$-functions tend asymptotically to
these constant values (if the $\beta$-functions for $\omega$-type of couplings are
considered). (Here we also suppressed the contribution from the anomalous
dimension $\eta$ since we know that this goes as some inverse powers
of the $\omega_{C}$ coupling and should be accurately included only
when higher-loop computation accuracy is required. We have $\eta\ll1$
and effectively we can take $\eta=0$, which is a judicious assumption
in the UV regime of the flow towards asymptotically free point.) As
a matter of fact, we see that the $\beta$-function of the $\omega_{C}$
is positive (with the precise value $\beta_{C}=\frac{199}{30}$), while
the $\beta$-function of the coupling $\omega_{{\rm GB}}$ is negative
(with the precise value $\beta_{{\rm GB}}=-\frac{87}{20}$). Since
in this regime, we reach an asymptotically free UV FP, where the $g$-couplings
go to zero (and correspondingly $\omega$-couplings tend to infinite
values), any RG flow towards such a FP should coalesce with the perturbative one-loop
RG flow for small values of the $g$-couplings.  We also see this as
a feature of our improved RG flows. Therefore, the locations of horizontal
asymptotes of the RG flow from the right (so for $k\to+\infty$) are universal and scheme-independent.

In the paper~\cite{JLK}, we studied the energy evolution of the running couplings towards the IR regime of the flow, concentrating on the IR fixed points of the system. The inclusion of threshold phenomena, which are present in any mass-dependent  renormalization, is of crucial importance in our analysis.
In fact, if we have studied only the simplified system of $\beta$-functions (\ref{simplsys}), we would not find any interesting behavior of the RG flow in the infrared (similarly to the case of QCD in the IR regime where the coupling grows stronger and gets out of the perturbative regime). We do not find any IR FP in such a simplified scheme. The $\beta$-functions from the system (\ref{simplsys}) for ${\cal X}={\rm const}$ are always constant, at any energy scale. To search for some non-trivial behavior in the IR, we must thus include some additional non-perturbative effects. This feature is brought about by our usage of FRG methods and account of decoupling of massive modes in the IR domain.

In order to find the FP's of the system in the IR regime, we must solve equations $\beta_C(k)=0\,$ and $\,\beta_{{\rm GB}}(k)=0$. One can see a big simplification here because in order to find zeros, we do not need to solve the full system~(\ref{71ac}). Actually, we can completely forget the denominators in ~(\ref{71ac}) and solve only Eqs.~(\ref{simplsys}), where threshold effects embodied in the factors $\Lambda/k^2$ are still taken into account. We also notice that the anomalous dimension $\eta$ does not influence the locations of the possible IR FP's, within the limits implied by our truncation ansatz used in FRG.

Numerical solutions of the equations $\beta_{C}(\kappa) =  0$ and $\beta_{{\rm GB}}(\kappa) = 0$ reveal that they are both satisfied at (approximately) simultaneous values of the rescaled energy scale $\kappa={k}/{\sqrt{|\Lambda|}}$.
This is a smoking gun for the fixed point of the RG flow.  These zeros are automatically zeros of the exact system~(\ref{71ac}).
Moreover, the location of the zeros is almost identical (up to 2\% accuracy)
for the couplings $\omega_C$ and $\omega_{{\rm GB}}$ for both cases of $\Lambda>0$ and $\Lambda<0$. The inclusion of higher-loop effects or extension of our truncation ansatz will make this agreement even stronger, so that in an exact fully non-perturbative theory,
the locations of two zeros coalesce into the one unique location of a genuine FP  for both couplings, in the infrared regime.

As we remarked above, the non-trivial form of the running arises because
we have included the effects of threshold phenomena. Let us, for definiteness,
analyze closer the case of the parameter $\Lambda>0$. It is straightforward
to understand the behavior of the exact $\beta$-functions $\beta_{C}$
and $\beta_{{\rm GB}}$ regarding their zeros and ensuing non-trivial
FP's in the IR domain. For this, it is important to find the behavior
of the MSS one-loop partition function treated as a rational function
of the energy scale $k$. The first pole/zero of the partition function
(when one is coming from large values of $k$) we find at $k^{2}=\frac{4}{3}\Lambda$,
and this is due to the factor $\det_{2T}\left(\hat{\square}-\frac{4}{3}\Lambda\hat{1\hspace{-1mm}{\rm I}}\right)$,
which is present in the denominator of the partition function (\ref{17ab}). This
latter implies that this factor appears in the FRG flow equation with the
positive coefficient, because it was a pole (not a zero) of the partition function.
As it was explained earlier, for the contribution of this factor to the Wetterich equation, we need to evaluate the following functional
trace
\begin{eqnarray}{\rm Tr}_{2T}\left(\frac{\left(\partial_{t}R_{k}-\eta R_{k}\right)\hat{1\hspace{-1mm}{\rm I}}}{\hat{\square}+R_{k}\hat{1\hspace{-1mm}{\rm I}}-\frac{4}{3}\Lambda\hat{1\hspace{-1mm}{\rm I}}}\right),\label{trexp}
\end{eqnarray}
which in the RG flow is with the coefficient $+1$. For details of the structure of the FRG flow equation the reader is referred to~\cite{JLK} and the formula (47) there. Here, we meet the
moment where we see the dependence on the cut-off kernel function $R_{k}=R_k(z)$,
so the dependence on the scheme of renormalization. We will now show that
for the qualitative features of the flow important for the existence
of non-trivial FP this dependence is immaterial. First, the trace in (\ref{trexp}) leads
to the expression in terms of the heat-kernel $B_{4}$ expansion coefficient of the operator,
namely to $\left(1-\frac{4}{3}\Lambda k^{-2}\right)^{-1}B_{4}\left(\hat{\square}_{2T}-\frac{4}{3}\Lambda\hat{1\hspace{-1mm}{\rm I}}_{2T}\right)$,
again with positive front coefficient. To get the front factor of the above expression we used the optimized (Litim) form of the cutoff kernel, $R_k(z)=(k^2-z)\theta(k^2-z)$~\cite{Litim2}, and we will remark on other possible choices below.  One can check that the $B_{4}$
coefficient of such an operator is also positive. Indeed, we find,
\begin{equation}
b_{4}\left(\hat{\square}_{2T}-\frac{4}{3}\Lambda\hat{1\hspace{-1mm}{\rm I}}_{2T}\right) \ = \ \frac{3}{5}\Lambda^{2}\, ,
\end{equation}
which is valid on MSS background and under the space volume integral (to produce the integrated $B_4$ coefficient).
This implies that in the flow of the action $\partial_{t}\Gamma_{k}^{L}$
(as evaluated on MSS background), we have the term $\left(1-\frac{4}{3}\Lambda k^{-2}\right)^{-1}\left(\frac{3}{5}\Lambda^{2}\right)$,
again with the positive sign. Hence, the expression for the $\beta$-function
$\beta_{{\rm GB}}$, which is the only one that can be read from the situation on MSS background, contains the factor $\frac{9}{40}\left(1-\frac{4}{3}\Lambda k^{-2}\right)^{-1}$,
when we again emphasize the positive sign.

By solving algebraically the linear system of the $\beta$-functions (\ref{simplsys}), we also
derive that in consequence the $\beta$-function $\beta_{C}$ contains
in turn the factor $\frac{9}{40}\left(1-\frac{4}{3}\Lambda k^{-2}\right)^{-1}$, but
with the minus sign. This opposite sign is the aftermath of the form
of the equations in the system: from MSS we derive only $\beta_{{\rm GB}}$
where the threshold factors reside, while on Ricci-flat background
we find only the combination $\beta_{C}+\beta_{{\rm GB}}$ without threshold
factors. Therefore, the threshold factors are inherited by the $\beta$-function
 $\beta_{C}$  solved algebraically but with the minus sign.
The appearance of the anomalous dimension $\eta$ (important for quantitative
description of the flow) does not change anything for what regards
the zero of the system of $\beta$-functions as we remarked before in \cite{JLK} (provided
that it is always negative $\eta\leqslant0$). The factor $\left(1-\frac{4}{3}\Lambda k^{-2}\right)^{-1}$
is the first one, which decides about the location of a vertical asymptote of the
$\beta$-functions $\beta_{C}$ and $\beta_{{\rm GB}}$, as seen on the plot of Fig.1. in \cite{JLK}, when the flow
comes from the UV direction. In general, about the positions of vertical asymptotes of the flow  decide the factors in the partition function (both in the denominator and in the numerator).

Now, one can look at this factor from the broader perspective. It
describes the decoupling (due to threshold phenomena) of the massive
modes with the mass square parameter on MSS background given by $\frac{4}{3}\Lambda$. The coefficient
in front is related to the $B_{4}$ coefficient of the corresponding
operator and that is why it is positive. In a general renormalization
scheme, this coefficient is also positive and the threshold factor
$\left(1-\frac{4}{3}\Lambda k^{-2}\right)^{-1}$ which shows the pole,
precisely at $k^{2}=\frac{4}{3}\Lambda$, must be present in this
form or in another more general one, but still in a form exhibiting the pole in the same place.
This is due to the gauge-invariant and universal fact that on MSS
background in Weyl conformal gravity we find stable (non-tachyonic)
$2T$ modes (spin-2 traceless) with the mass square given by $\frac{4}{3}\Lambda$.
And this is the highest mass square parameter in the spectrum of all
modes there. In a general framework, the general threshold factor could
look like this
\begin{eqnarray}
af\left(\left(1-\frac{4}{3}\Lambda k^{-2}\right)^{-1}\right)\, ,
\end{eqnarray}
where $a$ is a positive constant and $f(x)$ is some smooth, regular and positive
function at $x\geqslant0$. For the $\beta$-function $\beta_{C}$ this threshold
factor appears as
\begin{eqnarray}
bf\left(\left(1-\frac{4}{3}\Lambda k^{-2}\right)^{-1}\right)\, ,
\end{eqnarray}
with a constant $b<0$. This form manifests all universal features that we have discussed
above. Now, it is a matter of simple analysis of functions, that for
the {\rm GB} term coupling $\omega_{{\rm GB}}$, if $a>0$ and
$\beta_{{\rm GB}}{\to}-\frac{87}{20}<0$, when $k\to+\infty$, then
the running $\beta$-function $\beta_{{\rm GB}}(k)$ must meet a zero
for some $\sqrt{\frac{4}{3}\Lambda}<k_{{E}}<+\infty,$ because
$a\cdot\left(1-\frac{4}{3}\Lambda k^{-2}\right)^{-1}{\to}+\infty$, when $k\to\sqrt{\frac{4}{3}\Lambda}^{\,+}$. In the latter the  ``$+$'' superscript signifies that we approach the respective value from the above.
Similarly, for the Weyl term coupling $\omega_{C}$, if $b<0$ and
$\beta_{C}{\to}\frac{199}{30}>0$, when $k\to+\infty$, then the
running $\beta$-function $\beta_{C}(k)$ must meet a zero for some $\sqrt{\frac{4}{3}\Lambda}<k_{C}<+\infty,$
because $b\cdot\left(1-\frac{4}{3}\Lambda k^{-2}\right)^{-1}{\to}-\infty$, when $k\to\sqrt{\frac{4}{3}\Lambda}^{\,+}$.
These are the invariant features of the flow and that is why the IR
FP (for $k=k_{{{\rm GB}}}$ for the $\omega_{{\rm GB}}$ coupling and
for $k=k_{C}$ for the $\omega_{C}$ coupling) is universally present
in any renormalization scheme that aims at properly taking into account
threshold phenomena. It is obvious that any mass-dependent scheme comes with its own form of the function $f=f(x)$, positive for $x\geqslant0$, to supply the correct decoupling of heavy, massive modes in the IR regime. (As a such renormalization scheme we cannot, for example, select a DIMREG because it is mass-independent.)

 As we have argued, the form of the function $f(x)$  is irrelevant and the IR TP is scheme-independent. Mathematically, the zeros of threshold factors which appear in the expressions for the FRG-improved $\beta$-functions are the results of  the continuity of the function $f(x)$ and of the existence of the first vertical asymptote at $k^2=\frac43\Lambda$, and of the way how it is approached by $\beta$-functions. It is also of crucial importance that the factor $\left(k^2-\frac43\Lambda\right)$ first appeared in the denominator of the partition function (\ref{17ab}), so the sign was favorable to enforce the change of the sign of the $\beta_{{\rm GB}}$, when moving from negative universal one-loop value in the UV towards lower energies. This happens also because the vertical asymptote is approached from the right (higher energies) to $+\infty$. The change of the sign of $\beta_{{\rm GB}}$ must occur between two regimes: UV (when $\beta_{{\rm GB}}<0$), and IR near the first vertical asymptote (when $\beta_{{\rm GB}}\to+\infty$), so the zero line must be crossed for some energy scale $k=k_{{\rm GB}}$. A similar conclusion holds for the $\beta_C$ with inherited threshold phenomena obtained from the system (\ref{betasys2}). Moreover, it is natural to expect that $k_{{{\rm GB}}}\approx k_{C}$
since the difference can be associated only to higher-loop accuracy
error as one can easily see by comparing the numerical values, which are quoted below for $\Lambda>0$:
\begin{equation}k_{C} \ \approx \ 1.17709\sqrt{\Lambda}\quad{\rm and}\quad k_{{\rm GB}}\ \approx \ 1.19163\sqrt{\Lambda}\,.
\end{equation}
Finally,
one sees that the values of the energy characterizing the IR TP are
very close to the lower bound given by $k=\sqrt{\frac{4}{3}\Lambda}\approx1.1547\sqrt{\Lambda}$.
This means that we must inevitably find a FP in the infrared regime.
Therefore, the existence of the IR TP is a universal feature of the
exact (improved) RG flow, while the details of its location, slopes, and speeds of approaching the FP, etc., depend
on the particular choice of the renormalization scheme (or in FRG
terminology on the choice of the cut-off kernel function $R_{k}(z)$).

\subsection{Non-analyticity near turning point}
\label{s43}

Let us now observe that the system of $\beta$-functions Eqs.~(\ref{71ac}) and (\ref{Xeqn}) is not autonomous because the equations
depend on the initial conditions of the flow, not only on the actual
values of the couplings. In other words, we can see that these flow equations depend explicitly on the RG-time $t$ parameter, or on its exponential version $k=k_0e^t$. In the autonomous system, which is the case, for instance, for one loop in QED, in the dimensional regularization scheme or in simple momentum subtraction renormalization scheme, one has that $\beta=\beta(\omega)$ are functions of the actual values of couplings only. When
one includes threshold phenomena for massive modes, then the autonomy
of the system of RG flow equations is typically lost. This can be observed
also in our case. This conclusion is based on the comparison of
the RG running of the same $\beta$-functions, obtained for different initial
conditions of the flow. We see that even in the situation where the
actual values of the couplings are identical the corresponding
$\beta$-functions for two such flows are unequal. Viewed differently, we can observe
that the same values of the $\beta$-functions are attained for different
values of the actual couplings $\omega$, so the parameter
$t$ must also enter into dependence of the $\beta$-functions.

The lack of autonomy of the system of $\beta$-functions is the main obstacle against the possibility to express the $\beta$-functions in terms of couplings only. We remark that in more standard applications of FRG, FP's of the RG flow are looked for such systems $\beta=\beta(\omega)$ and conditions for FP's are conditions on the values of the couplings attained at the FP. In our case, for a genuine IR FP, we must have the additional condition that $t\to-\infty$. As explained in \cite{JLK}, we were able to find a TP for some finite energy scale and a true IR FP for any value of the couplings $\omega_C$ and $\omega_{{\rm GB}}$. Therefore, in our case we do not have any condition on the couplings at the FP --- they are fully unconstrained, or if analyzed in the space of all possible couplings, we have found not a single-point FP, but a 2-dimensional surface of FP's of the RG flow.

Here, we present the analysis near the turning point of the RG flow, where $\beta=0$, which occurs at $t=t_{*}$
and with the value of the coupling which is $\omega=\omega_{*}$ (this value
of the coupling $\omega_{*}$ depends on the initial condition of
the flow, that is on $\omega_{0}=\omega(t_{0})$ and $t_{0}$, while
$t_{*}$ is independent of them). We analyze the general situation for one representative coupling, but we can think of the TP as found in the system of $\omega_C$, $\omega_{{\rm GB}}$ couplings for $\kappa_*\approx1.18$. In the linear Taylor approximation, due
to the regularity of the $\beta$-function at the zero point understood as a function of the $t$ variable only, we can generically write
$\beta=\mathcal{A}(t-t_{*})$  for a constant coefficient $\mathcal{A}$ and this leads to the equation
\begin{eqnarray}
\frac{d\omega}{dt} \ = \ \mathcal{A}(t-t_{*})\, ,
\end{eqnarray}
which is solved by
\begin{eqnarray}
\omega-\omega_{*} \ = \ \frac{\mathcal{A}}{2}\left(t-t_{*}\right)^{2}\,,
\label{solnltp}
\end{eqnarray}
with the initial condition of ODE that $\omega(t_*)=\omega_*$. The above solution
 shows that there is a minimum of the coupling for the
value $\omega_{*}$ at $t=t_{*}$ for $\mathcal{A}>0$ (it is a local maximum when the constant coefficient $\mathcal{A}<0$). The case $\mathcal{A}>0$ we meet for the $\omega_{C}$
coupling, while for $\omega_{{\rm GB}}$ we have the opposite behavior (due to (\ref{betasys2})), so $\mathcal{A}<0$ there. For definiteness, here we consider the case $\mathcal{A}>0$. Inverting the relation in (\ref{solnltp}), we get that $t-t_{*}\sim\sqrt{\omega-\omega_{*}}$,
so the linearized $\beta$-function in terms of the $\omega$-coupling has the non-analytic form $\beta=\sqrt{2\mathcal{A}(\omega-\omega_{*})}$.
This signifies that there are two branches of the couplings (before
and after the minimum of the $\omega$ coupling is reached). Due to the square root involved, at TP, there is
a cusp-singularity in the local expression $\beta=\beta(\omega)$. The RG flow of the runnning coupling $\omega=\omega(t)$ has here a turning point because when $t$ is decreased the coupling $\omega$ first decreases and reaches a local minimum, to finally start growing again and going through the same values of $\omega$ for $t<t_*$. The $\beta$-function as function of $t$ smoothly crosses the zero line, while $\beta=\beta(\omega)$ has the cusp-behavior at $\omega=\omega_*$ and the double-valued behavior for $\omega>\omega_*$ with two (initially perfectly) locally symmetric branches of $\beta$ with opposite signs. One could also consider the stability
matrix of the linearized RG flow here at the TP.  One of its eigenvalue could be finite $\sqrt{2\mathcal{A}}$,
but $\frac{\partial\beta}{\partial\omega}$ at $\omega=\omega_{*}$
is formally positive infinity  on the upper branch,  so strictly speaking
there are no eigenvalues of the linearized RG flow around this point. On the lower branch, we have the opposite situation with formal negative infinity due to the existence of the cusp. This signals the failure of the linearization of the flow $\beta=\beta(\omega)$ around such a point. The flow is non-analytic at $t=t_*$.

Actually, for the $\beta$-function of the {\rm GB} term, the coefficient $\mathcal{A}<0$
because the zero of $\beta_{{\rm GB}}$ is reached from the other
side, so instead
of the local minimum for the coupling $\omega_{C}$, here there is a local maximum
for the coupling $\omega_{{\rm GB}}$. Moreover,  there is again formally a negative/positive infinite eigenvalue of the RG flow (depending on which branch one is moving on), hence the stability matrix cannot be properly defined. When we consider two couplings at the same time at TP, a question arises which coupling (only one) has to be chosen to locally eliminate the $t$ variable from the system of $\beta$-functions. We decided to remove $t$ in favor of the $\omega_C$ coupling. If the $\beta$-function of the {\rm GB} term, at the common TP (placed conventionally at $\kappa=\kappa_C$), is analyzed as a function of $\omega_{C}$,
not of the additional RG-time $t$ variable, and not of the $\omega_{{\rm GB}}$
coupling, then there is a finite positive off-diagonal value of the
linearized RG flow matrix because $\beta_{{\rm GB}}\sim\omega_{C}-\omega_{C*}$ with some finite coefficient. This is due to the facts that $\kappa_{{\rm GB}}>\kappa_C$ and that $\beta_{{\rm GB}}(\kappa_C)>0$.
Here, at the TP, we have a relation that $\omega_{C}-\omega_{C*}$ is proportional
to $(t-t_{*})^2$ with a positive coefficient from (\ref{solnltp}), hence then the two eigenvalues of the system
are zero for the {\rm GB} coupling and formally positive infinity for the $\omega_{C}$ coupling, and these two couplings formally are exact  eigenvectors of the matrix of the RG flow. Again, even in the case of two couplings the stability matrix cannot be determined.

The reader can easily see that the analysis presented above hinges on the fact that the location of the TP of the RG flow in RG-time coordinate $t_*$ is finite. If formally $t_*\to\pm\infty$, then the initial assumption about the local behavior of the $\beta$-function $\beta=\mathcal{A}(t-t_*)$ does not make any sense. Hence, the solution in (\ref{solnltp}) is not realized in this form. As it is known from the general theory of true FP's of the RG flows, they may only appear at the abstract theoretical RG-scale coordinate $t_*\to\pm\infty$. Then there are various theoretical ways how the asymptotics of the function $\beta(t)\to0$ is realized, but in most of the cases, it is possible to linearize the RG flow near the FP, when the $\beta$-functions are expressed entirely via the $\omega$-couplings. At such FP we generically find that $\beta(\omega)=\mathcal{A}(\omega-\omega_*)$ to the first infinitesimal level. Therefore, the flow is analytic and can be linearized to obtain finite derivative $\frac{\partial\beta}{\partial\omega}$ (in the case when we have many couplings, this is a linear matrix) giving us the stability coefficient. The form of the solutions $\omega=\omega(t)$ near the true FP's is typically exponential in $t$ variable and moreover they do not depend on the particular initial values of the flow $\omega_0=\omega(t_0)$, hence the system of the $\beta$-functions in the UV/IR becomes effectively autonomous. All the above statements about FP's are equivalent to each other and they show a clear distinction from the charasteristic of the RG flow at the stop of the flow which happens at TP's.

Also in the case of TP's the statements about the finiteness of $t_*$,  the linearity of $\beta=\beta(t)$ to the first order in $t-t_*$, the non-analyticity and the square-root-like singularity of the function $\beta=\beta(\omega)$, the lack of autonomy of the system of $\beta$-functions, and finally the impossibility to linearize the RG flow near TP and to define the stability matrix there are equivalent. For their derivation we have not used any additional assumption and this is the reason why in the proof of these equivalences one can easily go both ways. For example, from the non-analytic form of the $\beta$-function $\beta=\sqrt{2\mathcal{A}(\omega-\omega_{*})}$, one derives the local behavior of the $\beta$-function near the TP in $t$ variables: $\beta=\mathcal{A}(t-t_{*})$, which makes sense only for $t_*$ finite. That is why all the characteristics of the TP as the special point of the RG flow are tightly related and it differs from a true FP of RG. Similarly, from the AdS/CFT point of view true FP's of RG (both in IR or in UV) correspond to AdS backgrounds in asymptotic conformal regions of the gravitationally dual bulk spacetime, while the TP's correspond to bounce solutions holding in intermediate finite regions of spacetime characterized by some finite values of the AdS-like radial coordinate.

Following the above distinction between TP's and FP's, in the paper~\cite{JLK}, we continued the search for true IR FP's. For this purpose, we used the established fact that at $\kappa\approx1.18$ we found a common TP of the RG system. We exploited the infinitesimal form of the flow at TP and analytically extended it beyond the TP. We treated the TP as a good point from which we could start a new perturbation calculus driven towards the IR regime. Assuming perturbativity (in different couplings than in the UV FP), we were eventually able to find a genuine IR FP at $t\to-\infty$. We characterized this FP as non-trivial and non-Gaussian and computed the characteristic values of the couplings there, so called $\omega_{C**}$ and $\omega_{{\rm GB}**}$, which revealed to be non-vanishing. Moreover, we found that this IR FP is stable for both  perturbation directions given by the couplings $\omega_C$  and $\omega_{{\rm GB}}$. The IR-stable true IR FP is the main result to be used for further cosmological and conformal symmetry breakdown related applications of the QWG theory, see also Section \ref{5a}.

\subsection{Extension of the RG analysis to two-loop order}

In this subsection, we attempt to give a preliminary analysis
of the RG system of running coupling parameters in QWG, at the two-loop level. We base
our considerations only on algebraic, dimensional-analytic, and combinatorial arguments since a detailed computation of UV-divergences
and $\beta$-functions at this level is still beyond our computational capabilities. We assume that
numbers, we are dealing below with, are generic and they do not vanish, and we discuss the general structure of
the RG system. In particular, we touched upon the issue of the ``new'' $\beta$-function $\beta_{R}$, which is expected to be generated first time at the two-loop level. We analyze its suppression compared to other $\beta$-functions in the system and establish the hierarchy of them. Moreover, we also look at the universality properties of the $\beta$-functions for all three couplings $\omega_C$, $\omega_{{\rm GB}}$ and new $\omega_{R}$ at this level and in this way we strengthten and extend the well known results from the one-loop quantum level to QWG theory at two loops.

As remarked, in the paper \cite{JLK}, the first paper to deal with the divergences issue at the two-loop level in QWG, was the one by Fradkin and Tseytlin from 1984 \cite{Fradkin6}. However, the computation presented there is not complete, since only a subset of two-loop diagrams is analyzed.  Nevertheless, we agree with the authors' conclusion that it is very probable that the $R^2$ divergence shows up for the first time at the two-loop accuracy. This is in contradiction to the conjecture of `t Hooft and Mannheim \cite{Mannheim:2017}, who instead expect that the conformal symmetry on the quantum level is so powerful that this non-conformal $\beta$-function $\beta_{R}$ is vanishing to all orders and also non-perturbatively. This would be true, if the conformality was fully present at the quantum level (not only at the one-loop level, where it forces $\beta_{R}=0$). We do not think to be so, in accordance with \cite{Fradkin6, Duff}, because of the presence of conformal anomaly, already at the one-loop perturbative level. More issues related to the conformal anomaly we discuss in the special discussion section \ref{tanomaly}.

\subsubsection{Two-loop suppression of $\beta$-functions}
%
Let us remind that the one-loop action for QWG reads
\begin{eqnarray}
S \ =\!\int\!d^{4}x\sqrt{|g|}\left(\omega_{C}C^{2} \ + \ \omega_{{\rm GB}}\mathcal{G}\right)\,.\label{sactiontree}
\end{eqnarray}
This also served us as the truncation ansatz for the effective action $\Gamma$ that we used for the FRG computation. We re-emphasize that the term $R^2$ with the coupling $\omega_{R^2}$ is consistently not included at the tree-level and in the one-loop motivated RG flow equation, because such a term is \emph{not} generated by any quantum correction at the one-loop level.
In the original one-loop computation by Fradkin and Tseytlin \cite{Tseytlin}, the quantities
which are assumed to be small are $
\frac{1}{\omega_{C}}$ and $\frac{1}{\omega_{{\rm GB}}}\, .$
The loop expansion is precisely in  these quantities, that is at the
one-loop level we have
\begin{eqnarray}
\beta_{C}^{(1)} \ = \ \beta_{C}^{{\rm FT}}\quad{\rm and}\quad\beta_{{\rm GB}}^{(1)} \ = \ \beta_{{\rm GB}}^{{\rm FT}}\, ,
\end{eqnarray}
where the coefficients $\beta_{C}^{{\rm FT}}$, $\beta_{{\rm GB}}^{{\rm FT}}$
are simple numbers (\ref{betace}), while up to the two-loop order we must find
\begin{eqnarray}
\beta_{C}^{(2)} \ = \ \beta_{C}^{{\rm FT}} \ + \ a_{C,C}^{(2)}\frac{1}{\omega_{C}} \ + \ a_{C,{\rm GB}}^{(2)}\frac{1}{\omega_{{\rm GB}}}\, ,\label{twoloop1}
\end{eqnarray}
and
\begin{eqnarray}
\beta_{{\rm GB}}^{(2)} \ =\ \beta_{{\rm GB}}^{{\rm FT}} \ + \ a_{{\rm GB},C}^{(2)}\frac{1}{\omega_{C}}\ + \ a_{{\rm GB},{\rm GB}}^{(2)}\frac{1}{\omega_{{\rm GB}}}\, ,\label{twoloop2}
\end{eqnarray}
where the numerical coefficients $a_{C,C}^{(2)}$, $a_{C,{\rm GB}}^{(2)}$, $a_{{\rm GB},C}^{(2)}$, and
$a_{{\rm GB},{\rm GB}}^{(2)}$ are presently unknown, but it is certain that they do not depend on the couplings $\omega_C$, $\omega_{{\rm GB}}$. The two-loop form of the RG system presented above is the result of assuming the perturbative expansion in $
\frac{1}{\omega_{C}}$ and $\frac{1}{\omega_{{\rm GB}}}$ variables.

The UV-divergent part of the effective action at the one-loop level is given schematically by
\begin{eqnarray}
\Gamma^{(1)} \ = \ \!\int\!d^{4}x\sqrt{|g|}\left(\beta_{C}^{(1)}C^{2}\ + \ \beta_{{\rm GB}}^{(1)}\mathcal{G} \ + \ \beta_{R}^{(1)}R^{2}\right)\, ,\label{oneloopaction}
\end{eqnarray}
where we find that to one-loop accuracy we have
\begin{eqnarray}
\beta_{R}^{(1)}\ = \ \beta_{R}^{{\rm FT}}\ = \ 0\, ,
\end{eqnarray}
due to (partial) conformal symmetry still preserved at the quantum one-loop
level. This fact can be viewed as the one-loop remnant of full conformal symmetry present in the action at the tree-level (\ref{sactiontree}). At the two-loop level we expect $\beta_{R}^{(2)}$ not to vanish
and be given analogously by
\begin{eqnarray}
\beta_{R}^{(2)} \ = \ \beta_{R}^{{\rm FT}} \ + \ a_{R,C}^{(2)}\frac{1}{\omega_{C}} \ + \ a_{R,{\rm GB}}^{(2)}\frac{1}{\omega_{{\rm GB}}} \ = \ a_{R,C}^{(2)}\frac{1}{\omega_{C}}\ + \ a_{R,{\rm GB}}^{(2)}\frac{1}{\omega_{{\rm GB}}}\, ,\label{twoloop3}
\end{eqnarray}
where the coefficients $a_{R,C}^{(2)}$, $a_{R,{\rm GB}}^{(2)}$ are presently unknown numbers, whose non-vanishing (even of one of them), if unambiguously computed, would completely prove the conjecture of \cite{Fradkin6}. We explain that the possible structure term $a_{R,R}^{(2)}\omega_R^{-1}$ is not present since only $\omega_{C}$ and $\omega_{{\rm GB}}$ are the couplings in the original
(and also one-loop level) action. The coupling $\omega_{R}$ has to
be introduced (and renormalized) only from the two-loop level only. Actually, the reason for its introduction at the two-loop level action is the presence of the $R^2$ counterterm in $\Gamma^{(2)}$. We need to absorb such a covariant UV-divergent term and for this we need to include the $\omega_R R^2$ term in the bare action. This also means that for  perturbative computations at the level of three loops and higher, we must use the bare action (\ref{sactiontree}) corrected by the presence of this new term $\omega_R R^2$ with arbitrary coefficient $\omega_R$ (however, due to hierarchy and suppression of $\beta$-functions, as explained below, we should assume that its value is parametrically smaller than the values of other couplings $\omega_C$ and $\omega_{{\rm GB}}$ present in (\ref{sactiontree}), that is we should use $\omega_R\ll\omega_C,\omega_{{\rm GB}}$). For full one- and two-loop level quantum computations we can use the bare action as given in (\ref{sactiontree}) and this is reflected in the results for the $\beta$-functions to this accuracy given in (\ref{twoloop1}), (\ref{twoloop2}), and (\ref{twoloop3}). It is conceivable that if one wants to theoretically go to three-loop expressions for the $\beta$-functions of any of the coupling $X,Y=\omega_C$, $\omega_{{\rm GB}}$, or $\omega_R$, then the terms with the structure $a_{X,Y,R}^{(3)}\omega_{Y}^{-1}\omega_{R}^{-1}$ could appear in $\beta_X^{(3)}$ with non-vanishing coefficients $a_{X,Y,R}^{(3)}$.

The two-loop level UV-divergent part of the effective action takes then the form
\begin{eqnarray}
\Gamma^{(2)}\ =\!\int\!d^{4}x\sqrt{|g|}\left(\tilde{\beta}_{C}^{(2)}C^{2} \ + \ \tilde{\beta}_{{\rm GB}}^{(2)}{\cal G} \ + \ \tilde{\beta}_{R}^{(2)}R^{2}\right)\,,\label{twoloopaction}
\end{eqnarray}
so the new term $R^{2}$ is generated with the coefficient $\tilde{\beta}_{R}^{(2)}$,
which is always suppressed by one power of the small coupling $\frac{1}{\omega_{C}}$ or $\frac{1}{\omega_{{\rm GB}}}$, as in (\ref{twoloop3}).
Compared to the one-loop level action (\ref{oneloopaction}), where the counterterms were
multiplied by only numerical coefficients $\beta_{C}^{{\rm FT}}$ and
$\beta_{{\rm GB}}^{{\rm FT}}$ this is a suppression by additional power of small coupling. We conclude here, that the $\beta$-function $\beta_R$ when it finally shows up at the two-loop level is additionally suppressed with respect to other $\beta$-functions in the RG system. This signifies that the hierarchy of the $\beta$-functions is evident and the running of the $\omega_R$ coupling is very small, and that it was fully consistent to assume to the one-loop accuracy that $\omega_R=0$. This was the fact that we took advantage of in the truncation ansatz for $\Gamma$ that we used to model FRG to the one-loop level. Sincerely speaking, the significant $R^2$ term could be generated in the truncation ansatz $\Gamma$, but this does not happen immediately, and it requires a long RG-time since the running of $\omega_R$ is very slow. Our truncation ansatz for QWG is therefore internally consistent, at least in a big vicinity of the UV FP of RG. We just remark that we took care of the fact that at the two- and higher-loop level coefficients of UV-divergences are not the same as higher-loop $\beta$-functions of couplings (but they are in strict relations) and therefore we decorate the terms in (\ref{twoloopaction}) by additional primes.

If one uses the electric charge-like couplings defined for any coupling
$\omega$ (from the set $\omega_{C}$, $\omega_{{\rm GB}}$, and $\omega_{R}$) (cf., also the analysis in Subsection~\ref{UVAF})
by
\begin{eqnarray}
\omega_{i} \ = \ \frac{1}{g_{i}^{2}}\, ,\label{omdef}
\end{eqnarray}
then the corresponding $\beta$-function reads
\begin{eqnarray}
\beta_{i} \ = \ \beta_{g_{i}}\frac{d\omega_{i}}{dg_{i}} \ = \ -2g_{i}^{-3}\beta_{g_{i}}\, .
\end{eqnarray}
Consequently, we find that to the one-loop level accuracy
\begin{eqnarray}
\beta_{C}^{(1)} \ = \ -2g_{C}^{-3}\ \! \beta_{g_{C}}^{(1)}\, \quad {\rm and} \quad \beta_{{\rm GB}}^{(1)} \ = \ 2g_{{\rm GB}}^{-3}\ \! \beta_{g_{{\rm GB}}}^{(1)}\, ,
\end{eqnarray}
and hence $\beta_{g_{C}}^{(1)}\propto g_{C}^{3}\,$, $\beta_{g_{{\rm GB}}}^{(1)}\propto g_{{\rm GB}}^{3}\,$, and finally $
\beta_{g_{R}}^{(1)} \ = \ 0$.
Similarly, based on Eqs.~(\ref{twoloop1}) and (\ref{omdef}), up to the two-loop level we find
\begin{eqnarray}
\beta_{C}^{(2)} \ = \ -2g_{C}^{-3}\ \! \beta_{g_{C}}^{(2)} \ = \ \beta_{C}^{(1)}\ + \ a_{C,C}^{(2)}\ \! g_{C}^{2}\ + \ a_{C,{\rm GB}}^{(2)}\ \! g_{{\rm GB}}^{2}\, ,
\end{eqnarray}
which implies that
\begin{eqnarray}
\beta_{g_{C}}^{(2)} \ = \ -\frac{1}{2}g_{C}^{3}\ \! \left(\beta_{C}^{(1)} \ + \ a_{C,C}^{(2)}\ \! g_{C}^{2} \ + \ a_{C,{\rm GB}}^{(2)}\ \! g_{{\rm GB}}^{2}\right) \ = \ -\frac{1}{2}\left(g_{C}^{3}\ \! \beta_{C}^{(1)} \ + \ a_{C,C}^{(2)}\ \!g_{C}^{5}\ + \ a_{C,{\rm GB}}^{(2)}\ \! g_{C}^{3}\ \! g_{{\rm GB}}^{2}\right)\, .
\end{eqnarray}
We repeat verbatim for the {\rm GB} term coupling:
\begin{eqnarray}
\beta_{g_{{\rm GB}}}^{(2)} \ = \ -\frac{1}{2}\left(g_{{\rm GB}}^{3}\ \! \beta_{{\rm GB}}^{(1)}\ + \ a_{{\rm GB},C}^{(2)}\ \! g_{C}^{2}\ \! g_{{\rm GB}}^{3} \ + \ a_{{\rm GB},{\rm GB}}^{(2)}\ \! g_{{\rm GB}}^{5}\right)\, ,
\end{eqnarray}
and for the $\omega_R$ coupling (remembering that $\beta_{g_{R}}^{(1)} \ = \ 0$):
\begin{eqnarray}
\beta_{g_{R}}^{(2)} \ = \ -\frac{1}{2}\left(a_{R,C}^{(2)}\ \!g_{C}^{2}\ \!g_{R}^{3} \ + \ a_{R,{\rm GB}}^{(2)}\ \!g_{{\rm GB}}^{2}\ \!g_{R}^{3}\right)\ = \ \mathcal{O}\left(g^{5}\right)\, .
\end{eqnarray}
We see that at the \emph{leading order} the $\beta$-function $\beta_{g_{R}}$ is
proportional to the fifth power of the $g$-couplings.  Again, compared to the expressions for $\beta_{g_C}$ and $\beta_{g_{\rm {\rm GB}}}$, which to the leading order (which is a one-loop order) go like $g^3$, this is a two-loop suppression. We derive that this suppression is present independently of the form of the couplings used in QWG theory.

When using electric-like-couplings $g_i$, we are in a comfortable situation that perturbative calculus is conveniently done in these couplings (compare to a different case for $\omega$-like couplings). For example,
we write the one-loop effective action  $\Gamma^{(1)} \ \propto \  \mathcal{O}(g^{0})$ as a perturbation in small
couplings $g_{i}$ to the classical action $S \ \propto \  \mathcal{O}(g^{-2})$,
and then the genuine corrections at the two-loop level are
\begin{eqnarray}
{\Gamma}^{(2)} \ \propto \  \mathcal{O}(g^{2})\, .
\end{eqnarray}
The coupling of the $R^{2}$ term in the effective action $\Gamma$ for QWG is generated from the level of (at least) two loops and its first coefficient is proportional to $g^{2}$, so compared to other
terms like $C^{2}$ or $E$ present at the one-loop level and multiplied by
the coefficients of order $g^{0}$, this coefficient is again highly suppressed.
We also draw a parallel that analogously like in QWG considered above, in QCD, the $\beta$-functions scale like: $\beta^{(1)}_g\sim g^3$ for the one-loop level and $\beta^{(2)}_g\sim g^5$ for the two-loop level, and similarly the tree-level action scales as $S \ \propto \  \mathcal{O}(g^{-2})$, the one-loop level effective action scales as $\Gamma^{(1)} \ \propto \  \mathcal{O}(g^{0})$, and the two-loop effective action scales as $\Gamma^{(2)} \ \propto \  \mathcal{O}(g^{2})$, where $g$ is the Yang-Mills coupling parameter.

\subsubsection{Universality of two-loop order $\beta$-functions in Weyl square gravity}

Let us now present a general argument for the universal properties of the RG system of $\beta$-functions of all three couplings $\omega_C$, $\omega_{{\rm GB}}$, and $\omega_R$ in QWG at the second loop level.
Our exposition is loosely based on  Ref.~\cite{thooft}.
First, we recall that to the one-loop accuracy all $\beta$-functions of perturbative couplings are universal and do not depend on the choice of the renormalization scheme, and are also gauge-fixing independent, if the corresponding couplings are in front of dimension four gauge-invariant terms in the action. Of course, the form of the $\beta$-functions depend on the version of the couplings used (whether this is an $\omega$-type coupling, or $g$-like coupling, or something else) as clearly seen in the formulas from the last subsection. Here, we analyze the change of the $\beta$-functions under a general redefinition of the couplings. Such redefinition mimics the change done by using a different regularization or renormalization scheme, or the change of the gauge-fixing conditions. Provided that this change is continuously connected to the point $g_i=0$ and that it is analytic in the couplings $g_i$, we find that the type of the coupling is preserved. (We remark that the change between $\omega$-type and $g$-type couplings is not regular at the point $g=0$, hence it does not satisfy the above criterion, and as can be clearly seen from expressions in Eqs.~(\ref{betace}) and (\ref{betasys1loopg}), the corresponding $\beta$-functions are different, though they are obviously related to each other.) For the analysis in QWG, we also use the fact that $\omega_{{\rm GB}}$ coupling cannot appear in any Feynman diagram computation, since it multiplies the Gauss--Bonnet term, which is a total derivative in $d=4$. Hence, there cannot be any dependence on $\omega_{{\rm GB}}$ in any perturbative $\beta$-function of the theory. This observation simplifies also the analysis of the previous subsection, where we can effectively set everywhere $1/\omega_{{\rm GB}}\to0$.

The $\beta$-function of the $R^2$ term is expected first
to show up at the two-loop level and in the form
\begin{eqnarray}
\beta_{R}^{(2)} \ = \ \frac{d\omega_{R}}{dt} \ = \ \mathfrak{a}\frac{1}{\omega_{C}}\, ,
\end{eqnarray}
where $\mathfrak{a}\equiv a_{R,C}^{(2)}$ is some constant coefficient, which has not yet been computed. Effectively
we have here the system of two couplings $\omega_{R}$ and $\omega_{C}$ and three $\beta$-functions. (In accordance with the remark made above, we do not have $\omega_{{\rm GB}}$ as the coupling on which these  $\beta$-functions could depend.)
First, the $\beta$-function of the coupling $\omega_{C}$ is in its sector universal (it is
behaving effectively like the only coupling in the town). By looking at the formula in Eq.~(\ref{twoloop1}) when we set $a_{C,{\rm GB}}^{(2)}$ to zero, we see that this RG sector is identical with a two-loop sector of a theory with only one coupling $\omega_C$. As it is known from~\cite{thooft},  two-loop $\beta$-function for any QFT system of only one unique coupling is also invariant under the general coupling redefinition transformations.

Now, the question is only about the $\beta$-function $\beta_{R}$. We prove below that its two-loop value is also universal.
In the two-couplings system ($\omega_R$, $\omega_C$) we are allowed to change couplings only
in the following way, written to the leading order $\omega^0$,
\begin{eqnarray}
\omega'_{R} \ = \ \omega_{R} \ + \ A_{0}\ + \ A\frac{\omega_{R}}{\omega_{C}} \ + \ B\frac{\omega_{C}}{\omega_{R}} \ + \ \mathcal{O}\left(\omega^{-1}\right)\, ,\\
\omega'_{C} \ = \ \omega_{C} \ + \ C_{0}\ + \ C\frac{\omega_{R}}{\omega_{C}}\ + \ D\frac{\omega_{C}}{\omega_{R}}\ + \ \mathcal{O}\left(\omega^{-1}\right)\,.
\end{eqnarray}
But there exists a requirement that in any gauge choice or parametrization method,
the $\beta$-functions $\beta'_{R}$ and $\beta'_{C}$, when expressed in terms of primed couplings, cannot depend on
$\omega'_{R}$, because
there are no Feynman diagrams depending on this coupling (here on $\omega'_{R}$) needed to be considered to the two-loop level of accuracy. (The bare action to the two-loop level contains only one term $\omega_C C^2$.) Hence, one can check, that the only possible
changes of the couplings are shifts according to:
\begin{eqnarray}
\omega'_{R} \ = \ \omega_{R}\ + \ A_{0}\, ,\\
\omega'_{C}\ = \ \omega_{C} \ + \ C_{0}\, ,
\end{eqnarray}
and then, of course, we have that the transformed $\beta$-functions read
\begin{eqnarray}
&&\beta'_{R} \ = \ \beta_{R} \ = \ \mathfrak{a}\frac{1}{\omega_{C}} \ = \ \mathfrak{a}\frac{1}{\omega'_{C}} \ + \ \mathcal{O}\left(\frac{1}{\omega_{C}^{`2}}\right)\, ,\\[2mm]
&&\beta'_{C} \ = \ \beta_{C} \ = \ \beta_{C}^{\rm FT} \ + \ a_{C,C}\frac{1}{\omega_{C}} \ = \beta_{C}^{\rm FT} \ + \ a_{C,C}\frac{1}{\omega'_{C}}\ + \ \mathcal{O}\left(\frac{1}{\omega_{C}^{`2}}\right)\, ,
\end{eqnarray}
to the quadratic order in the coupling $\omega'_C$. This proves the universality in the $\omega_{R}$ and $\omega_{C}$
sector of the expressions for the two-loop $\beta$-functions of these couplings
of the theory.

Similarly, we find for the Gauss--Bonnet coupling $\omega_{{\rm GB}}$,
which does not appear in any perturbative Feynman rule of the theory,
so none of the $\beta$-function can depend on it.  Hence, the only permissible changes of couplings
are given by
\begin{eqnarray}
\omega'_{C} \ = \ \omega_{C} \ + \ C_{0}\, ,\\
\omega'_{{{\rm GB}}} \ = \ \omega_{{\rm GB}} \ + \ D_{0}\, .
\end{eqnarray}
Then, according to (\ref{twoloop2}), the $\beta$-function is to two-loop order accuracy
\begin{eqnarray}
\beta_{{\rm GB}}^{(2)} \ = \ \beta_{{\rm GB}}^{\rm FT} \ + \ a_{{\rm GB},C}^{(2)}\frac{1}{\omega_{C}}\, ,
\end{eqnarray}
and a transformed $\beta$-function is
\begin{eqnarray}
\beta'_{{{\rm GB}}} \ = \ \beta_{{\rm GB}}^{\rm FT} \ + \ a_{{\rm GB},C}^{(2)}\frac{1}{\omega_{C}} \ = \ \beta_{{\rm GB}}^{\rm FT} \ + \ a_{{\rm GB},C}^{(2)}\frac{1}{\omega'_{C}} \ + \ \mathcal{O}\left(\frac{1}{\omega_{C}^{`2}}\right)\, ,
\end{eqnarray}
so again there is a universality of this  two-loop expression for $\beta_{{\rm GB}}^{(2)}$. Basically, all the couplings in QWG at the two-loop level, behave like if they were in separate one-coupling sectors of the couplings' space of the theory. Everything boils down to the fact that the RG system of $\beta$-functions at two-loop level depends only on one coupling $\omega_C$. This is a special feature of QWG, that to the two-loop level we need only to deal with one coupling of the Weyl square term, the {\rm GB} term will not have any impact perturbatively and only from the third loop we need to include the new coupling $\omega_R$ of the $R^2$ term. This new term $\omega_R R^2$ in the bare action is heralded by the presence of trace anomaly already at the one-loop level, to which discussion we turn now.

\section{Discussion of the trace anomaly issue
\label{tanomaly}}
%
Let us now discuss the issue of the trace (or conformal) anomaly. Firstly,we recall the following fact about trace anomaly in ordinary Yang-Mills gauge theories in $d=4$ spacetime dimensions. Standard YM theories are described by the action $ S_{g}=-\frac{1}{2}\!\int\!d^{4}x\,{\rm tr}(F_{\mu\nu}^{2})$. This entails that on the classical level the theory is conformally invariant and hence the trace anomaly on this level vanishes.  One could compute the trace of the classical energy-momentum tensor (obtained by Hilbert method of variation with respect to some fiducial metric tensor of any curved background) and then it is found to vanish in agreement with conformal symmetry. On the other hand, if the theory is not very special, then for a generic situation at the quantum (loop) level there is a non-vanishing $\beta$-function $\beta$ of the YM coupling, which is a signal of the presence of trace anomaly. Due to the RG-invariance in the effective action of the YM model at the one-loop level we have the term
\begin{equation}{\rm tr}\left(F_{\mu\nu}\log\left(\frac{D^2}{\mu^2}\right)F_{\mu\nu}\right),\label{termFsquareF}\end{equation}
where $D^2$ is the square of the gauge-covariant derivative operator and $\mu$ is this renormalization scale which was also used to put renormalization conditions for fields and compensate $\mu-$dependence of the running dimensionless YM coupling. This RG running is in an invariant manner (and universal to one-loop level) described by the non-vanishing $\beta$-function $\beta$.
One could ask for the conformal properties of the newly generated term (\ref{termFsquareF}) in the finite pieces of the effective action. Clearly, it is not even scale-invariant due to the presence of the dimensionful renormalization energy scale $\mu$. Hence this term cannot be conformally invariant. If one tries to evaluate the trace of the energy-momentum tensor coming from this term it is definitely non-zero, hence there is a  trace anomaly due to quantum effects. The trace was zero on the classical level (as the consequence of the presence and full realization of conformal symmetry), but it is non-zero on the quantum level, so there is an anomaly related to the fact that on quantum level there is no conformal symmetry anymore in this model. As it was elucidated above, this fact is in tight links with the presence of non-vanishing $\beta$-functions of the model. To one-loop level the value of the $\beta$-function is universal, but as  better observables we could choose scattering amplitudes in this model. While they are constrained by scale-invariance on the tree-level, due to the presence of the term (\ref{termFsquareF}) at one-loop level the scattering amplitudes show behavior which explicitly breaks scale-invariance (and conformal-invariance in particular too). Simply, the amplitudes depend on the energy scale, while at tree-level they do not. For example, the 4-gluon scattering amplitude on tree-level is expressed only by the square of the constant value of the YM coupling at tree-level and there is no any dependence on the incoming gluon energies. After inclusion of the finite corrections, like the term (\ref{termFsquareF}), in the effective action, the resulting 4-gluon scattering amplitudes do show dependence on energies, as a consequence of the presence of the scale $\mu$ in the term (\ref{termFsquareF}). These are the observable results and we just interpret them that the scale-invariance is not present on the quantum level of this model and does not constrain amplitudes anymore.
Simply saying, quantum physics of standard pure gauge theories is more complicated and more interesting than just what was on the tree-level constrained by scale-invariance. The amplitudes are more complicated and show more intricate behavior with energy scales as the result of freedom from conformal symmetry. The trace anomaly is not here a problem and can be used to generate some terms in the effective action. Actually, it signals that quantum gauge field theory are more interesting and worth studying. Here conformal symmetry was never meant to be used as local gauge symmetry hence its lack after inclusion of quantum corrections (due to polarization effects of gluons) is not problematic. We could say that conformal symmetry  was an accidental symmetry of the tree-level (classical) generic YM theory and at the quantum level we have seen that it is not there anymore. Since we do not put much of emphasis to this symmetry in gauge field theories, then its loss it is not a big deal like this happens with other accidental symmetries. We can sacrifice easily this conformal invariance and we shall not regret it (although now the computations on the quantum level are much more involved). This is the physics of YM theories, and still it is fully consistent without conformal invariance on the quantum level.

There is a completely different attitude for quantum conformal gravity theories, since there by definition we want to use the conformal symmetry with its fundamental and not accidental role. This fundamental role is signified in that we want to use this symmetry to constrain the form of all possible terms in the gravitational action (only $C^2$ action is acceptable in $d=4$ dimensions), to constrain the spectrum of the model, which then is different from a generic spectrum of four-derivative gravitational theory. These things shall not be understood as accidents due to conformal symmetry, they are definite virtues. And the fact that they happen is not surprising but rather demanded and expected from the fundamental role of conformal symmetry in constraining gravitational interactions. We want to use full conformal group as the gauge symmetry group of conformal-gravitational interactions. By gauging full 15-parameter conformal group we give the decisive and fundamental role to both the local Poincar\'{e} and local conformal symmetries. Only together they could be used to put a  control over quantum theory of gravitational interactions.

Let us now return to the issue of the trace (or conformal) anomaly.
For the energy-momentum tensor of the total system (matter + gravity) we define the trace in the classical and quantum cases respectively as
\begin{eqnarray}
T \ = \ g_{\mu\nu}T^{\mu\nu}\quad {\rm or} \quad \langle T \rangle \ = \ \langle g_{\mu\nu}\hat T^{\mu\nu} \rangle\, .
\end{eqnarray}
For quantum conformally invariant theories we should find that $\langle g_{\mu\nu}\hat T^{\mu\nu} \rangle=0$.
By explicit computation on the classical level for conformal models we find that $T=0$,
but on the quantum level
 one can find $\langle T\rangle \neq0$, namely by general arguments~\cite{Bonora:1985cq}
\begin{eqnarray}
\langle T\rangle \ = c C^{2}\ + \ a\mathcal{G} \ + \ a_{P}P \ + \ a_L \Box R\, ,
\end{eqnarray}
where $P=\epsilon_{\mu\nu\lambda\zeta}R^{\mu\nu}{}_{\rho\sigma}R^{\rho\sigma\lambda\zeta}$ is the parity-odd Pontryagin density, which can be a priori excluded in parity-preserving theories (as QWG is), and the last term is ambiguous as it depends on the renormalization prescription. Coefficients $a$ and $c$ are the central charges that can be directly related to the corresponding $\beta$-functions studied in the previous subsections (namely $c$ and $a$ are gauge-dependent for spin $3/2$ and $2$ but $c+a$ is not~\cite{Christensen:1978gi,Christensen:1978md}). One can find a straightforward relation between the logarithmically divergent part of the quantum effective action and the anomalous terms above. In particular, at one loop one can check that
\begin{eqnarray}
\langle T\rangle \ = \ \frac1{(4\pi)^2}b_4\, ,
\end{eqnarray}
so that $c=\beta^{(1)}_C$ and $a=\beta^{(1)}_{{\rm GB}}$.
In our case this anomaly poses a problem: the classical theory (before quantization) was with local conformal symmetry,
while quantum theory is without it.
Quantum effects break the gauge symmetry of the model 
on the full quantum level (e.g., conformal analogues of Slavnov--Taylor identities are not satisfied).
Consequently, the symmetry is without any power to constrain UV-divergences, Green functions, the form of the quantum corrections to the effective action or scattering amplitudes.
So, in order to get a better grasp on the trace anomaly issue it is imperative to use the effective action $\Gamma$ rather than semiclassical arguments.

First, in pure $C^2$ gravity we note that the $\beta$-function of Weyl coupling $\beta_C$ is non-vanishing at the quantum level (starting from first loop).
Hence in the effective action $\Gamma_{\rm fin}$ we must have a term
\begin{eqnarray}
\beta_C C \log \left(\frac{\square}{\mu^2} \right)C\, ,
\end{eqnarray}
with $\mu$ being an arbitrary renormalization scale.
This is due to RG-invariance of the total effective action $\Gamma$ and the $\beta$-function $\beta_C$.
In conclusion the total action in $d=4$ is
\begin{eqnarray}
\Gamma_{\rm tot} \ = \  \Gamma_{C^2}+\Gamma_{\rm eff} \supset \sqrt{g}\left[C^2+\beta_C C \log \left(\frac{\square}{\mu^2} \right)C\right]\, .
\end{eqnarray}
Relevant observation pertaining to it is that $\Gamma_{\rm tot}$ is not conformally invariant action in $d=4$!
Definitely, conformal symmetry is not present on the quantum level in Weyl square gravity.

There is the perturbative trace anomaly issue, but the problem is also at non-perturbative level. This means that the theory despite being perturbatively renormalizable is however, non-perturbatively non-renormalizable~\cite{Duff}. This is because the radiative correction break Weyl symmetry in a dramatic way. Pure $C^2$ gravity is anomalous.
In other words, there is RG running (towards IR) as explained in this work, but the problem is at high energies (UV-limit) for overall consistency of the model of conformal gravity. It is very important to cancel this anomaly since conformal symmetry appears in a local (gauged) version. Similarly, like in gauge theory models we have to ascertain that there are no gauge anomalies due to fermions in the matter sector since the presence of such anomaly would be disastrous for the gauge symmetry on the quantum level. Theory would be inconsistent again because of lack of symmetry on the quantum level. In a  perturbative vein there would be pop up perturbative UV-divergences  that we will not be able to absorb in any form of counterterms possible in gauge-invariant actions. These counterterms would not be gauge-invariant since on the quantum level there is no any symmetry constraining their form. And if there is not a constraining (or forbidding) symmetry law on the quantum level in QFT everything what is possible to be generated is generated (according to the QFT's Murphy's law).

The possible trace anomaly resolution is given below. It is known that there exists special matter content coupled to $C^2$ gravity such that $\beta_C=0$ (and all other $\beta$-functions in the matter sector vanish too) and then the terms $C\log\square C$ in the effective action $\Gamma$ are not generated at all. For example, in the case of ${\cal N}=4$ supergravity  Fradkin and Tseytlin \cite{Fradkin6} showed this can be achieved by coupling it to ${\cal N}=4$ super-YM (SYM) gauge field theory model, which is known to be conformal on flat spacetime. This is brought by example of coupling ${\cal N}=4$ supergravity due to Fradkin and Tseytlin to ${\cal N}=4$ SYM  gauge field theory model which is known to be conformal on flat spacetime. Then the quantum conformality is present in the coupled model and the anomaly is cancelled due to mutual interactions between gravitational and SYM sectors. Both these sectors are necessary for the successful trace anomaly cancellation. Then if we have secured the presence of conformal symmetry on the quantum level we can exploit the possibility to constrain scattering amplitudes, correlation functions, form of the effective action $\Gamma$ etc. It is well known that conformal symmetry is omnipotent in constraining the form of the terms in the effective action $\Gamma$ because in a fixed dimensionality of spacetime only few terms (finite number of them!) are conformally invariant compared with an inifinite number of terms which are consistent with gauge symmetry or coordinate diffeomorphism invariance. The theory could be so powerful that the quantum effective action is idempotent with respect to quantization procedure (understood as going from classical tree-level action functional $S$ to the fully quantum effective action functional $\Gamma$).
Moreover, this also opens the possibility to resolve GR-like singularities using conformal symmetry of some exact solutions of  quantum $\Gamma$, which now are also exact solutions of the classical theory.

From a different perspective, if there is no conformal (gauged) anomaly
on the quantum level, then the theory is endowed with quantum scale-invariance,
there are no non-zero $\beta$-functions, no RG flow and the theory sits at the UV FP of RG. This is a situation that has to be secured in UV by embedding the quantum Weyl gravity theory in the bigger picture, e.g., in the ${\cal N}=4$ supergravity or in a twistor superstring theory of Berkovits and Witten. Another possibility is that the theory reaches in the UV regime a non-trivial (non-Gaussian) FP of RG. With the additional conditions that the dimension of the critical surface on which putative UV FP lies, is finite, this realizes the Weinberg's Asymptotic Safety scenario (AS). In these circumstances, UV-divergence problem is avoided and there are no $\beta$-functions (even on the non-perturbative level) since the theory sits at the FP. There is no RG flow and scale-invariance of the situation is enhanced to full conformal invariance and the UV-action that describes such a FP is an action of conformal field theory (CFT). There is a whole discipline of studies of CFT's but quite a little is known about CFT when the gravity is quantum and dynamical. But this is a desired CFT, that describes the quantum theory of conformal gravitational interactions at very high energies. As with any other CFT to fully describe it we would need to give the set of primary conformal operators and their anomalous dimensions (conformal weights). This constitutes the set of CFT data. Based on it we could describe any correlation function within this gravitational CFT. To get away from the FP and start RG flow at lower energies some deformation operator must be added to such CFT. That is, we think that by deforming CFT by adding an operator which is not conformally invariant, we start a non-trivial RG flow away from the UV FP. And then following the flow we can reach even the domain of low-energy physics.

As it was pointed out long ago, the pure $C^2$ theory may reveal to be non-renormalizable from the level of two loops onward. This is due to the presence of Weyl anomaly~\cite{Duff}.  Therefore some authors claim that because of the non-zero trace anomaly $C^2$ gravity does not survive quantization and hence it is inconsistent on the quantum level till the conformal anomaly (CA) is made vanish. Its most direct effect may be the presence of the $R^2$ term in the divergent part of the effective action. (Such term does not show up at one-loop, but it is expected at the two-loop level because all symmetries that could forbid its presence there are not realized on the quantum level). The UV-divergences which could be covariantly collected in the term $R^2$ are not absorbable by the counterterms of the original Weyl theory and that is why the theory may show up to be perturbatively non-renormalizable, when a definite answer will be given to the presence or not of the $R^2$ term on the two-loop level~\cite{Fradkin6}.

We emphasized that in our work~\cite{JLK}, we worked in the fixed-dimension regularization scheme in $d=4$ because this should not produce any spurious (e.g., dimension dependent) Weyl-symmetry violating terms in the effective action. So, we confined ourselves  to the cutoff scheme at fixed dimensionality of spacetime. In this way we do not discard or  avoid the trace anomaly (which indeed shows up already at the one-loop level), and the statement of its existence is scheme-independent. Strictly speaking, the onset of the trace anomaly is renormalization prescription independent because the $\beta$-function is zero only at the fixed point and the existence of the fixed point is indeed renormalization prescription independent. On the other hand, the actual non-zero value of the trace of the energy-momentum (i.e., trace anomaly) is renormalization prescription dependent, since the $\beta$-function is universal only to the second order in the coupling constant. The one-loop value is scheme-independent but from two loops on we expect that such dependence will start to develop. The conformal anomaly is proportional to the $\beta$-functions of the model. On different vein, these $\beta$-functions can be read off from perturbative UV-divergences of the effective action, hence the problem of trace anomaly is a problem rooted in the UV regime or, in other words, in an UV-completion of the theory.

At the UV FP which is asymptotically free for pure $C^2$ gravity the conformal anomaly is still  there. The $\beta$-functions for omega-type couplings are not vanishing at this FP (couplings multiplying directly invariants like $C^2$ or $\mathcal{G}$). Only $\beta$-functions in  alpha-type couplings (similar to electric charge-like couplings) vanish but the anomaly in terms of $\omega_C$  and $\omega_{{\rm GB}}$ is non-zero even in the UV FP.

Let us summarize here our approach towards the quantization of $C^2$ theory and RG running towards the IR limit. We start in the UV fixed point where theory is exactly Weyl-invariant (just a single dimensionless coupling) and the renormalization group flow deforms the theory (i.e., induces other terms) as the theory flows towards IR fixed point. It turns out that the theory has a non-trivial (non-Gaussian) fixed point in IR and hence it is asymptotically safe in IR or, in another words, it is non-perturbatively renormalizable in IR. This asymptotic safety (AS) is in accordance with Weinberg definition~\cite{Weinberg}, but this time for the IR, rather than UV sector. Our situation with the issue of trace anomaly is slightly different since we have asymptotic safety in IR. So we do not assume that theory is valid at all energy scales  down to IR. It very quickly flows to the IR FP that is so close to the UV FP that the $R^2$ term does not even have enough (RG-)time to appear in the effective action (it does not appear there at one-loop level), and after IR FP is reached we do not have anymore scale invariance. In other words, we surpass the problem of perturbative non-renormalizability by non-perturbative  renormalizability in the IR regime.

One may wonder whether following RG trajectory  towards lower energy one ends up at a reasonable point (be it Gaussian-like or Banks--Zaks FP) or  diverge. In the latter case, the theory would need some kind of IR-completion or protection against infrared  problems and IR-divergences. For this, if we have AS in IR, then the theory is non-perturbatively renormalizable and thus solves all such problems near IR FP.
%
%

As already mentioned, the problems related to conformal anomaly are essentially related to the UV regime.  
One might thus guess that having AS in IR FP does not a priori helps in solving UV problems.
For this,  we need to provide evidence for a separate UV FP. And it is obvious that looking locally in the  parameter space the existence of IR FP does not imply anything for the existence of any UV FP.
One possible scenario that can allow to infer possible existence of UV FP for QWG is to embed the theory in a broader  context of more fundamental (a better behaved) theories like twistor string theory or ${\cal N}=4$ conformal supergravity. In particular,  that the bosonic $C^2$ sector that we consider can be understand as a truncation of the anomaly-free ${\cal N}=4$ supergravity of Fradkin and  Tseytlin~\cite{Tseytlin}. The super ${\cal N}=4$ case of Fradkin and Tsetylin was later seen to provide the exceptional solution to the issue of trace anomaly~\cite{Tseytlin} (however, it seems that there is still a unitarity problem for this model). It is the latter theory which is in the UV without any problem, there is no conformal anomaly, no non-trivial $\beta$-functions. This theory is perturbatively renormalizable and UV-finite and hence there are no any problem with quantization.
%
%
In our previous work~\cite{JLK} we consider the IR version of QWG when we reduce and consistently integrate out degrees of freedom and we study only the bosonic spin-2 sector. Or in a different perspective, QWG appears as the low energy limit of twistor string theory  where there are not apparent problems with conformal anomaly.  Simply the conformal anomaly is a problem of the UV regime but we  assume that somehow this problem is solved there and we start with the theory, which is consistently reduced, in the intermediate  energies.
%
We assume that such UV FP exists and somehow addresses the anomaly issue, and then we  try to see if such a postulate is consistent.  At the same time there is a subtle difference with what  some people typically mean by UV problem of conformal gravity. We believe that most  practitioners just start perturbatively from IR FP (hence assume Gaussian or Banks--Zaks IR FP) and deduce the  problem in UV by increasing energies in the RG flows. Natural question arise, what about if one cannot even start perturbatively from IR as in our case (and also, e.g., in QCD).
What then one can say about UV FP?
We offer a plausible scenario that is logically consistent and reasonably well  motivated. We assume that UV FP is just a full-fledged critical point in a series of hypothetical phase transitions that the Universe has undergone in its very early stage.  It might be FP that descends from the spontaneous symmetry breaking (SSB) phase of twistor string theory or ${\cal N}=4$ conformal supergravity. In any case SSB FP has exact conformal symmetry and the simplest theory in $d=4$ that lives at that critical manifold is QWG (only one  coupling $\omega_C$).
Then we may hypothesize about UV-completion in a form  of string theory or something similar but clearly we do not have to do this now, if we have some strong evidences for UV FP and for solving some UV problems there. Moreover, if someone would like to see the RG evolution from IR FP towards UV, then he would probably need to go through horrible technical problems that would be even more complicated than in QCD.   We are more like physicists, who  start with asympotically free (deconfined) QCD and then deduce non-perturbative (confining) IR  phase.  This way is far simpler than going vice versa.

It should be stressed that when using FRG we do not a priori assume that there is any IR FP. This is a clear advantage of FRG approach. We just chose the truncation ansatz, which in our case is motivated by perturbative one-loop results. Should we have known the higher-loop (or even better the full non-perturbative) effective action then we could utilize its form and set up more informed truncation ansatz, which in turn could provide more reliable (or more refined) results.

\section{Physics beyond IR critical point
\label{5a}}


In this section we study the physics of the 2nd order phase transition that is associated with the IR FP of QWG discussed in Section~\ref{4sc}. The characteristic (in fact defining) feature of 2nd order phase transitions is existence of the order parameter field  that has zero vacuum expectation value (VEV) in the symmetric phase and non-zero VEV in the broken phase. Typical method used in this context is theory of effective potentials. This approach will be presented in the following two subsection. In Subsection~\ref{reflection} we will see that one can harness the conformal anomaly to say something more on the structure of an actual critical point and corresponding phases in its vicinity.


\subsection{Composite Hubbard--Stratonovich field as an order-parameter field, emergent Starobinsky's model
\label{5.1.a}}

Because we require that our theory should induce Einstein action in the low-energy limit we rewrite the $R^2$-part in the Bach action (\ref{PA2}) with the help of the Hubbard--Stratonovich transformation~\cite{expjizba,Hubbard,Stratonovich}   as
\begin{eqnarray}
\exp(iS_{R^2}) \ \equiv \ \exp\left(\frac{i}{6 \alpha^2} \int d^4 x \ \! \sqrt{|g|} \ \! R^2\right)\ = \ \int {\cal D} \sigma \ \! \exp\left[-i \!\!\int d^4 x \ \! \sqrt{|g|} \ \!\left(\frac{\sigma R}{2} \ \! +  \ \! \frac{3}{32 \ \! \omega_C} \ \!\sigma^2\right) \right]\!.
\label{PA2P}
\end{eqnarray}
It is clear that the HS transformation (\ref{PA2P}) is nothing but a simple identity based on a functional Gaussian integral.
Although the auxiliary HS field $\sigma(x)$ does not have a bare kinetic term, one might expect that due  to loop   corrections  the renormalized action  will  develop  in
the IR regime a gradient term, which then allows
to identify the HS boson with a bona fide propagating mode.
This mechanism is  well-known  from condensed matter theory~\cite{Sachdev,AS}
and  particle physics~\cite{MS1,MS2,Col,KL3}.
A typical example is  obtained when
the BCS superconductivity is reduced to its low-energy
effective level. There the HS boson coincides with the disordered
field whose dynamics is described via the Ginzburg--Landau
equation~\cite{GL}.
%
%

The $\sigma$ field in (\ref{PA2P}) can be separated into a background field $\bar{\sigma} \equiv \langle{\sigma}\rangle $ corresponding to VEV of $\sigma$ plus fluctuations $\delta\sigma$. Since  $\bar{\sigma}$ is dimensionful it must be zero in the case when the
theory is (globally) Weyl-invariant. On the other hand, when the Weyl symmetry is broken, $\bar{\sigma}$ develops a non-zero value. So, the $\sigma$ field  plays the role of the order-parameter field.
%
%
With the benefit of hindsight we further introduce an {arbitrary} (hyperbolic) mixing angle
$\vartheta \in {\mathbb{R}}$ and make the following splitting
\begin{eqnarray}
S_{R^2} \ = \  S_{R^2} \cosh^2\! \vartheta  \ - \  S_{R^2} \sinh^2 \!\vartheta \, .
\end{eqnarray}
If we now apply the HS transformation only to the $S_{R^2} \sinh^2 \!\vartheta $ part of the action (\ref{PA2}), we obtain
%
\begin{eqnarray}
S_{R^2} \ = \
\int \!d^4x\ \! \sqrt{|g|} \ \! \left[ -
\frac{\sigma}{2} \ \! R
\  +  \ \frac{2\omega_C  \cosh^2\! \vartheta}{3} \ \!
 R^2
\ + \  \frac{3}{32 \omega_C  \sinh^2 \!\vartheta}\ \!
\sigma^2\right]\, ,
\label{3rdac}\end{eqnarray}
and hence the full action (\ref{PA1}) can be rewritten (modulo topological term)  as
\begin{eqnarray}
S \ = \ \int \!d^4x\ \! \sqrt{|g|} \ \! \left[ - 2\omega_C R_{\mu\nu}^2 \ - \
\frac{\sigma}{2} R
\  +  \ \frac{2\omega_C  \cosh^2\! \vartheta}{3} \ \!
 R^2
\ + \  \frac{3}{32 \omega_C  \sinh^2 \!\vartheta}\ \!
\sigma^2\right]\, .
\label{3rdacd}
\end{eqnarray}
%
By employing the Weyl symmetry one can formally generate a gradient term for the $\sigma$ field already on the level of a bare-action. In fact, if we perform the Weyl rescaling of the form ${g}_{\mu\nu} = |\sigma|^{-1} \tilde{g}_{\mu\nu}$  we get the gradient term
\begin{eqnarray}
-\sqrt{|\tilde{g}|} \ \! \frac{3}{2} \ \! \frac{\sigma \Box \sigma}{\sigma^2} \ + \ \ldots\, ,
\end{eqnarray}
(dots refer to higher order derivative terms of the $\sigma$ field). It should be, however, clear that such a gradient term  cannot define
genuine propagating mode since it depends on the conformal scaling and hence it is a gauge dependent concept. This is actually consistent with the fact that the WG has no propagating scalar degree of freedom on the bare-action level.
On the other hand, when the scale symmetry is broken then the $\sigma$-field will be trapped in a particular broken phase with specific kinetic as well as potential term, cf. Eq.~(\ref{a.37bb}).

Although the full theory described by the action $S$ is manifestly independent of the mixing angle
$\vartheta$, truncation of the perturbation series at a finite loop order will inevitably destroy this independence.
The optimal result can be is reached via the principle of minimal sensitivity~\cite{Stev,Bender2,KS-F}, which posits that
if a perturbation theory depends on some unphysical parameter (e.g.,  $\sinh^2 \vartheta$ in our case) the best result is achieved if each perturbation order has the weakest possible dependence on the parameter $\vartheta$. As a result, the value of $\vartheta$ is determined at each loop order
from the vanishing of the corresponding derivative of effective action.

Let us now see that the VEV $\bar{\sigma}$ indeed acquires a non-zero value at low enough energies. Since the corresponding phase transition must be of 2nd order ($\bar{\sigma}$ must change smoothly from zero in symmetric phase to non-zero in broken phase) this is tantamount to showing that the IR FP must exist and that it is associated with the (global) scale-symmetry breakdown.
To this end we employ the standard effective-action methodology in the mean field approximation. In particular, we replace in (\ref{3rdacd}) all $\sigma$ fields with their VEV $\bar{\sigma}$ and compute the effective potential by integrating over metric-field fluctuations. This can be  done again in the framework of the York decomposition discussed in Section~\ref{quantization}. In order to simplify our computations we use as the background the Minkowski flat spacetime. This in turn means that our computations serve only as proof of the principle that one should expect
a phase transition in QWG with interesting cosmological implications in its broken phase. For more precise cosmological analysis one should, of course, use a wider class of Bach-flat backgrounds.

As discussed in~\cite{expjizba}, the one loop-effective potential acquires the form
\begin{eqnarray}
V_{\rm{eff}}(\bar{\sigma},\vartheta) &=& - \frac{9}{4^4\pi^2\omega_C^2} \frac{ \bar{\sigma}^2}{(4 \sinh^2 \vartheta + 1)^2}\left[\log\left(\frac{3 \bar{\sigma}}{2 \omega_C (4 \sinh^2 \vartheta + 1)\mu^2}   \right) \ - \ \frac{3}{2}  \right] \nonumber \\[2mm]
&&+ \ \frac{3  \bar{\sigma}^2}{4^3 \pi^2 \omega_C^2} \left[\log\left(\frac{\bar{\sigma} }{2 \omega_C \mu^2}\right) \ - \ \frac{3}{2}\right] \ - \ \frac{3  \bar{\sigma}^2}{4^2 \omega_C \sinh^2 \vartheta }\, ,
\end{eqnarray}
and this result is true both in dimensional~\cite{expjizba} and $\zeta$-function~\cite{knap} renormalization scheme. Here $\mu$ is the renormalization scale.

The corresponding VEV $\bar{\sigma}$ is obtained through minimizing $V_{\rm{eff}}$ which gives
\begin{eqnarray}
\bar{\sigma}(\vartheta) \ = \ 2\omega_C \mu^2 \ \!\exp\left[\frac{3 \sinh^2\vartheta \log\left(\frac{3}{4\sinh^2\vartheta \ + \ 1}\right) \ + \ 16 \pi^2 \omega_C (4\sinh^2\vartheta \ + \ 1) }{\sinh^2\vartheta (32 \sinh^4\vartheta \ + \ 16\sinh^2\vartheta -1 )} \  + \  1 \right]\, .
\label{VEV}
\end{eqnarray}
We might note that the value $\bar{\sigma} = 0$ is {\em not} a local minimum for $V_{\rm{eff}}$ when $\sinh^2\vartheta > (\sqrt{6}-2)/8$ because in such a case $V_{\rm{eff}} <0$, while for $\bar{\sigma} = 0$ one has $V_{\rm{eff}} = 0$. In order to see whether situation with $\bar{\sigma} \neq 0$ can be realized we employ the principle of minimal sensitivity, namely we require that
\begin{eqnarray}
0 \ = \ \frac{dV_{\rm{eff}}(\bar{\sigma}(\vartheta),\vartheta) }{d \sinh^2 \vartheta} \ = \ \frac{\partial \bar{\sigma}(\vartheta)}{\partial \sinh^2 \vartheta}\frac{\partial V_{\rm{eff}}}{\partial \bar{\sigma}} \ + \ \frac{\partial V_{\rm{eff}}}{\partial \sinh^2 \vartheta} \ = \ \frac{\partial V_{\rm{eff}}}{\partial \sinh^2 \vartheta}\, .
\end{eqnarray}
This equation admits two branches of real solutions~\cite{expjizba}. The branch that corresponds to the symmetric phase ($\bar{\sigma} = 0$)
satisfies the equation $\sinh^2 \vartheta = 0.02592337 - 0.0000197 \alpha^2 + \mathcal{O}(\alpha^2)$. The broken-phase branch ($\bar{\sigma} \neq 0$) corresponds  to large values of $\sinh^2 \vartheta$ with actual value depending on $\omega_C$. Consequently,  we can rewrite to order $\mathcal{O}(1/\sinh^4 \vartheta)$  Eq.~(\ref{VEV}) as
\begin{eqnarray}
\bar{\sigma}(\vartheta) \ = \ 2\omega_C \mu^2 \ \!\exp\left[1 + \frac{8 \pi^2 \omega_C }{\sinh^2 \vartheta(\omega_C)} \right]\, .
\end{eqnarray}
In particular, for any value of the dimensionless coupling
$\omega_C$, we can choose the renormalization  scale
$\mu$, in such a way that $\bar{\sigma}\sim 1/\kappa^2$, which in turn will guarantee
phenomenologically correct gravitational forces at
long distances. Consequently, Newton's constant $\kappa^2$ is dynamically generated.
Owing to the last term in (\ref{3rdac})-(\ref{3rdacd}), the appearance of a cosmological constant in the low-energy limit of the broken phase is also consequence of QWG. In the following it will be convenient to rescale $\sigma \mapsto \sigma/\kappa^2$ so that $\bar{\sigma}\sim 1$ and the field itself is dimensionless.

As already mentioned in Section~\ref{quantization}, the local scale symmetry dictates that the scalar degree of freedom must decouple from the on-shell spectrum of QWG. When the conformal symmetry is broken the scalar field reappears in action through a radiatively induced gradient term of the HS field $\sigma$. The explicit form of the kinetic term can be decided from the momentum-dependent part of  the $\sigma$-field self-energy $\Sigma_{\sigma}$. In Ref.~\cite{expjizba} it was shown that the corresponding
leading order gradient term is of the form  $\frac{1}{2\kappa^2 \bar{\sigma}} \partial_{\mu} \sigma \partial^{\mu} \sigma$.
By assuming that in the broken phase a cosmologically relevant metric is that of the Friedmann--Robertson--Walker (FRW), then, modulo a topological term, the additional constraint
\begin{eqnarray}
\int d^4x \sqrt{|g|}  \ \!  3 R_{\mu\nu}^2 = \ \int d^4x \sqrt{|g|}  \ \!  R^{2}\, ,
\end{eqnarray}
holds due to a conformal flatness of the FRW metric~\cite{Birrel}.  It was argued in~\cite{expjizba} that from (\ref{PA2}) and (\ref{3rdac}) one obtains in the broken phase the steepest descent (or WKB)  gravitational action of the form
%
%
%
\begin{eqnarray}
&&\mbox{\hspace{-4mm}}S_{{b.p.,\sigma}} \ = \  -
\frac1{2\kappa^2}\int d^4x\, \sqrt{|g|}
\left(\sigma R - \xi^2 R^2 - \frac{(\partial_{\mu} \sigma)^2}{\bar{\sigma}} -
2\Lambda \sigma^2\right) \, ,
\label{a.37bb}
\end{eqnarray}
In the long-wave limit one can neglect fluctuations of the $\sigma$ field and consider that $\sigma$ is basically described by its VEV, i.e., $\sigma = \bar{\sigma}$. This yields in the broken-phase the mean-field effective action
\begin{eqnarray}
S_{{b.p.,\bar{\sigma}}} \ = \ - \frac{1}{2\kappa^2}\int d^4x \sqrt{|g|}  \ \! (R  \ - \  \xi^2 R^2 \ - \ 2\Lambda)\, ,
\label{star}
\end{eqnarray}
By comparing (\ref{star}) with (\ref{3rdacd}) we obtain that
%
%
\begin{eqnarray}
\Lambda  \ = \ \dfrac{3}{32\omega_C \kappa^2\sinh^2 \!\vartheta}, \quad  \xi^2 \ = \  \dfrac{4 \omega_C\kappa^2 \sinh^2 \!\vartheta}{3}\,  \;\;\;\ \Rightarrow \;\;\; \Lambda  \ = \  \frac{1}{8\xi^2}\, ,
\label{39.aa}
\end{eqnarray}

The action (\ref{star}) is nothing but the Starobinsky action  with the cosmological constant.
In the Starobinsky model (SM) the linear Einstein term determines the long-wavelength
behavior while the $R^2$-term dominates short distances
and drives inflation, which is followed by the
gravitational reheating with the decrease of $R^2$~\cite{Capozziello}.
In phenomenological cosmology, the SM represents metric gravity with a curvature-driven inflation.
In particular, it does not contain any fundamental scalar field that could play the role of an inflaton field, even though a scalar field/inflaton formally appears when transforming the SM to the Einstein frame~\cite{Whitt}.
%
We stress that the cosmological constant $\Lambda$ is entirely of a geometric origin (it descends from the QWG) and it enters with the opposite sign in comparison with the usual matter-sector induced (i.e., de Sitter) cosmological constant. In the following we will call such gravitation-induced cosmological constant as {\em gravi-cosmological} constant
and it represents what we call the {\em dark side of QWG}.

In this connection the following point deserves mentioning.
Since the conformal symmetry prohibits the existence of a (scale-full) cosmological
constant, the gravi-cosmological constant must correspond to a scale at which the conformal
symmetry breaks, which in turn determines the cut-off scale of the $\sigma$-field.
The magnitude of $\xi$ in the SM is closely linked to the scale of
inflation~\cite{Duff}.
Using the values relevant for the Cosmic Microwave Background radiation (CMB) with $50-60$ $e$-foldings, the Planck data~\cite{Planck2,Planck3} require  that $\xi\sim 10^{-13}$GeV$^{-1}$
or equivalently $\xi/\kappa\sim 10^5$.
The vacuum energy density is thus $\rho_{\Lambda} \equiv \Lambda/\kappa^2
\sim 10^{-10}(10^{18} \mbox{GeV})^4$, which corresponds to a zero-point energy density of a
scalaron with an ultraviolet cut-off at about $10^{15}-10^{16}$GeV. This coincides with a range
of the GUT inflationary scale.
For compatibility with an inflationary-induced large structure formation the conformal symmetry should be broken before (or during) inflation. We will further discuss this issue in the following section.

%
%

\subsection{Broken phase QWG and hybrid inflation}

By now it is well recognized  that the scalar cosmological perturbations are (nearly)
scale-invariant with the value of the spectral index $n_s =0.97$, which is tantalizingly close to 1 --- exact scale invariance~\cite{expjizba,Mannheim:2011is}. The combination of data from Planck and BICEP2/Keck Array BK15
~\cite{Planck3} tightens the upper bound on the tensor-to-scalar ratio to obtain:
$r  < 0.056$ at 95$\%$ CL. These include, e.g.,
the SM, the non-minimally coupled model
($\propto \phi^2 R/2$) with a $V(\phi)\propto \phi^4$-potential,
inflation model based on a Higgs field and the so-called universal attractor models~\cite{Planck3}.

In the previous subsection we have seen that the long-wave limit of QWG approaches in the broken phase the SM.
It might be thus interesting to see what inflationary cosmology can be deduced from the action (\ref{a.37bb}).
In particular, we would like to see to what extend the predictions of CMB perturbations based on the Lagrangian~(\ref{a.37bb}) are compatible with Planck and BICEP2/Keck Array BK15 cosmological data.

To this end one can set up for
$S_{{b.p.,\sigma}}$ a dual description
in terms of a non-minimally coupled auxiliary
scalar field $\lambda$  with the action~\cite{Whitt,Capozziello}
\begin{eqnarray}
S_{\{\sigma,\lambda\},J}  &=&  - \frac{1}{\kappa^2}\int d^4 x
\, \sqrt{|g|}\left(\frac{\sigma + 2\xi \lambda}{2}\ \!R  + \frac{\lambda^2}{2}
- \frac{(\partial_{\mu} \sigma)^2}{2\bar{\sigma}} - \Lambda
\sigma^2\right).\label{jordan.fr.1}
\end{eqnarray}
This is basically a HS-transformed $S_{b.p.,\sigma}$ with $\lambda$ being the HS-field.
To analyze (\ref{jordan.fr.1}) we choose to switch from the Jordan frame to the Einstein frame~\cite{Capozziello}
where the curvature $R$ enters without a non-minimally coupled fields $\sigma$ and $\lambda$. This is obtained via
rescaling: $g_{\mu\nu}\ \mapsto (\sigma + 2\xi {\lambda})^{-1}g_{\mu\nu}$,
giving
\begin{eqnarray}
\mbox{\hspace{-8mm}}S_{\{\sigma,\lambda\},E}  &=&  - \frac{1}{\kappa^2}\int d^4 x \ \! \sqrt{|g|}\left[\frac{\tilde{R}}{2} -
\frac{3\xi^2 (\partial_{\mu}\lambda)^2 }{(\sigma +2\xi \lambda)^2} -\frac{3\xi (\partial_{\mu}\lambda) (\partial^{\mu}\sigma)}{(\sigma +2\xi \lambda)^2}  - \frac{(\partial_{\mu}\sigma)^2  }{2\bar{\lambda}(\sigma +2\xi \lambda)} \nonumber \right.\\[2mm]
&-& \left. \frac{3 (\partial_{\mu}\sigma)^2}{4(\sigma +2\xi \lambda)^2}  + \ \!\frac{\lambda^2}{2(\sigma +2\xi \lambda)^2} - \frac{\Lambda \sigma^2  }{(\sigma +2\xi \lambda)^2}\right].
\label{einst.fr.1}
\end{eqnarray}
The above metric rescaling is valid only for  $(\sigma + 2\xi {\lambda})>0$.
The action (\ref{einst.fr.1}) can be brought into a diagonal form if we pass from fields $\{\sigma, \lambda\}$ to $\{\sigma, \psi\}$ where the  new field  $\psi$ is obtained via the redefinition
$\lambda = [\exp(\sqrt{2/3} |\psi|) -\sigma]/(2\xi)$.
In terms of  $\psi$ the action reads
\begin{eqnarray}
&&\mbox{\hspace{-6mm}}S_{\psi, E} =  - \frac{1}{\kappa^2}\int d^4 x \sqrt{|g|}\left[\frac{\tilde{R}}{2} -
\frac{1}{2}(\partial_{\mu}\psi)^2  + V(\psi,\sigma)
\  -e^{-\sqrt{2/3}|\psi|}\ \!\frac{(\partial_{\mu}\sigma)^2}{2\bar{\sigma}}\right],
\label{einst.fr.2}
\end{eqnarray}
where $V(\psi,\sigma) = \frac{1}{8\xi^2}\left( 1- 2\sigma e^{-\sqrt{2/3}|\psi|}\right)$. The strength of $\sigma$-field oscillations is controlled by the size of a coefficient in front of the $\sigma$-gradient term, i.e.,  $e^{-\sqrt{2/3}|\psi|}/\kappa^2$. In particular, the local squared amplitude of the $\sigma$-field fluctuations is of order
\begin{eqnarray}
\delta \sigma^2(x) \ = \  \langle (\sigma(x) - \bar{\sigma})^2\rangle \ \sim \ \kappa^2 e^{\sqrt{2/3}|\psi(x)|}\, .
\end{eqnarray}
%
At large values of the dimensionless scalar field $\psi$, i.e., at values of the dimensionful field $\tilde{\psi} = \psi/\kappa$ that are large compared to the Planck scale, the gradient coefficient is very small and $\sigma$-field rapidly fluctuates. Assuming that QWG was broken before the beginning of inflation, then after a brief period of fierce oscillations the $\sigma$-fluctuations are strongly damped at around the value ${\psi} \lesssim 10 $ (i.e., $\tilde{\psi} \lesssim 10 m_p$). Indeed, if the fluctuations  $\sqrt{\delta \sigma^2(x)}$  are smaller that $10^{-17}$, i.e., smaller than the GUT inflationary scale, then
\begin{eqnarray}
\kappa \exp\left(\frac{1}{2}\sqrt{\frac{2}{3}}|\psi(x)|\right) \ \lesssim \ 10^{-17}    \;\;\; \Rightarrow \;\;\; \tilde{\psi} \ \lesssim \  10 m_p\, .
\end{eqnarray}
From that period on, the $\sigma$-field settles at its potential minimum at $\bar{\sigma} =1$.
Note that $V(\psi,\bar{\sigma})\leq 1/(8\xi^2) \ll m_p^2$, which is a necessary condition for a successful inflation.  At values of $\tilde{\psi} \sim 10 m_p$, the potential $V(\psi,\bar{\sigma})$ is sufficiently flat to produce the phenomenologically acceptable slow-roll inflation, with the inflaton field $\psi$. From the inflationary potential in the Einstein frame we can infer that this case represents
hybrid inflation, in the sense that the inflationary potential is due to non-zero $\bar{\sigma}$.
Using the slow-roll parameters
\begin{eqnarray}
&&\mbox{\hspace{-11mm}}\epsilon \ = \ \frac{1}{2}m_p^2 \left(\frac{\partial_{\psi}V(\psi,\bar{\sigma})}{V(\psi,\bar{\sigma})} \right)^2, \;\;\;\; \eta \ = \ m_p^2 \ \!\frac{\partial^2_{\psi}V(\psi,\bar{\sigma})}{V(\psi,\bar{\sigma})}\, ,
\end{eqnarray}
($\partial_{\psi} \equiv \partial/\partial\psi$) one can write down the tensor-to-scalar ratio $r$ and the spectral index $n_s$ in the slow-role approximation as~\cite{Planck3}
\begin{eqnarray}
r \ = \ 16  \epsilon, \;\;\;\; n_s \ = \ 1- 6\epsilon + 2\eta\, .
\end{eqnarray}
In terms of the number $N$ of $e$-folds left to the end of inflation
\begin{eqnarray}
N \ = \ - \kappa^2 \int_{\psi}^{\psi_f} \!\!\! d\psi\ \!  \frac{V(\psi,\bar{\sigma})}{\partial_{\psi}V(\psi,\bar{\sigma})} \ \approx \ \frac{3}{4\bar{\sigma}} e^{\sqrt{2/3}|\psi|} \, ,
\end{eqnarray}
($\psi_{f}$ represents the values of the inflaton at the end of inflation, i.e., when $e^{-\sqrt{2/3}|\psi|}\sim 1$) one gets
\begin{eqnarray}
&&\mbox{\hspace{-10mm}}  n_s \ \approx \ 1- \frac{2}{N},\;\;\;\; r \ \approx \ \frac{12}{N^2}\, ,
\end{eqnarray}
which for $N = 50 \div 60$ (i.e., values relevant for the CMB) is remarkably consistent with Planck data~\cite{Planck2,Planck3}.

While during the inflation, the $\sigma$-field is constant (due to a large coefficient in front of the gradient term) allowing a {\it large-valued} inflaton field  descend slowly from potential plateau,
inflation ends gradually when  $\sigma$ regains its canonical kinetic term, and a {\it small-valued} inflaton field picks up kinetic energy.
%
From  (\ref{einst.fr.2}) the dominant interaction channel at small $|\psi|$  is  $(\partial_{\mu}\sigma)^2 |\psi|$, hence the vacuum energy density stored in the
inflaton field is transferred to the $\sigma$ field via inflaton
decay $\psi \rightarrow \sigma + \sigma$ (reheating), possibly preceded by a non-perturbative stage (preheating).


Note also, that the gravi-cosmological constant $\Lambda$ that was instrumental in setting the inflaton potential in (\ref{einst.fr.2})
has the {\it opposite} sign when compared with
ordinary (matter-sector induced) cosmological constant.

Inflationary scheme discussed can be naturally incorporated in a broader theoretical context of ``conformal inflation'' model, which has been the subject of much recent investigation~\cite{Linde,Costa,Turok,Vanzo1}. Setting the inflation period around the time of Weyl-symmetry breakdown brings about a number of attractive features, including: natural justification for inflation models with a plateau, technically accessible inflationary correlators, specific predictions for the B-mode vorticity fluctuations in CMB powerspectra, etc.
Let us also notice that the existence of a single scalar field with cutoff at around the GUT scale and coupled to broken phase of QWG  (e.g., our $\sigma$ or GUT Higgs field)  would contribute with a {\it positive} zero-point energy that could offset gravi-cosmological $\Lambda$ and leave behind a small observable cosmological constant.
This could provide a viable mechanism for breaking down 120 orders of magnitude difference that presently exists between
theoretical predictions and astronomical observations of $\Lambda$. Consequently, the dark side of QWG would not be ultimately so dark.
\subsection{Some reflections on anomaly matching \label{reflection}}

We have seen that the effective potential approach provides a powerful method for discussing physics in both broken and unbroken phase.  Unfortunately, the physics in the immediate vicinity of the critical point
is poorly grasped by this approach. The point is that due to the appearance of long-range correlations, perturbative approaches are quite unreliable. It might seem rather surprising that the physics in the vicinity of the critical point might be for CFT's discussed in terms of conformal anomaly. The reason for this is basically twofold. First, conformal anomaly has an universal structure phrased in terms of two central charges (see Section~\ref{tanomaly}). Second, there is a powerful theorem~\cite{Schwimmer:2010za} that ensures that there is an anomaly matching across the critical point of the 2nd order phase transition.

In Section~\ref{4sc} we have considered a RG flow of QWG, which ends up in an IR FP that is associated to a CFT. So one could wonder how this IR FP should be associated to low-energy gravity where the scale symmetry is absent. One intriguing possibility is that the IR CFT has a moduli space of vacua, such that in all of them but the one at the ``origin'' conformal (or scale) invariance is spontaneously broken. Such examples are known, expecially in superconformal field theories~\cite{Fayet:1978ig}. This allows to identify a broken phase of the theory where the analytic structure of the amplitudes is not constrained by conformal invariance, but still the numerical value of the trace anomalies can be possibly matched (as proven, e.g., in Ref.~\cite{Schwimmer:2010za} for CFT's with global conformal invariance).
This poses strong constraints on the broken phase theory that is in spirit very similarly to what happens in {\em 't Hooft anomaly matching}~\cite{hooftanomaly}. In particular, requiring for the two cental charges
\begin{equation}
c_{<} \ = \ c_{>}\,,\quad  \;\;\; a_{<} \ = \ a_{>}\,,
\end{equation}
(``$>$'' refers to {\em physical} energy scales above the FP in question and ``$<$'' refers to {\em physical} energy scales below the FP~\footnote{ Here we should stress that {\em physical scale} is not the same as the {\em running scale} in the RG flow. The running RG-flow scale is always re-parameterized so that it reaches its limiting value ($\pm \infty$) at the FP. In physics the critical point that is represented by a RG FP always happens at some finite energy scale --- phase transition scale.}) it was shown to be possible to determine the couplings of the action for the dilaton which emerges in the broken phase as the Goldstone boson related to the spontaneously broken conformal symmetry~\cite{Schwimmer:2010za}. Actually, in a supersymmetric scenario where the trace anomaly is in the same supermultiplet as $R$-symmetry charge this boils down to the usual 't Hoft anomaly matching condition for which the generating functional of amplitudes accounting for anomalies is well understood. This could make it possible to understand the IR regime in the broken phase
in a quite novel way. We leave this interesting and important topic for our future investigation.

\section{Conclusions\label{concl}}


In this paper, based on the approach of functional renormalization group, we have mainly investigated the IR physics of the quantum Weyl gravity, which turns out to be surprisingly rich and interesting, in particular in connection with the structure of its IR FP and related cosmological implications.


Important point that we have utilised here, was the fact that at one loop the $R^2$-infinities of the effective action are absent so that we do not observe the appearance of the anomalous conformal mode.  Although it is expected that such situation will change at two loops (contrary to chiral anomalies, which are one-loop exact), one can still view this fact as an indication of a specific feature of the model that could potentially persist even to higher loops, where a conclusive calculations are still missing. In our FRG approach the IR FP (for the two involved couplings) develops already at the one-loop level and hence the prospective observational consequences of the trace anomaly do not take over before the Weyl symmetry is dynamically broken.

Problem that typically besets higher-derivative gravity theories is the unitarity issue. For QWG this is commonly phrased in terms of spin-$2$ ghost field. As already mentioned in the introduction there exist various remedies to this issue in the literature.
If unresolved, the ghost problem prevents QWG from being a possible UV completion of Einstein's general relativity.
But this conclusion is based solely on the analysis of a classical Hamiltonian or on the structure of tree-level propagators. Basically for any physical process that explicitly manifests the ghost problem, there is usually an implicit assumption that the perturbative analysis reflects the true physical spectrum. It is quite instructive to compare QWG with QCD. The latter is also renormalizable and asymptotically free theory but in this case one is accustomed to the fact that the physical spectrum bears no resemblance to the perturbative degrees of freedom. In particular, the gluon is not in the physical spectrum. This is a consequence of confinement, but it is also understood more directly in terms of the behavior of the full gluon propagator. Essentially there is an IR suppression of the propagator that is sufficient to remove the gluon pole. Similarly, we have seen that the FRG analysis of QWG shows that the IR FP is non-Gaussian hence QWG theory is non-perturbatively renormalizable in IR.

The presence of ghosts in the linearized spectrum  and the quantum breaking of conformal invariance (trace anomaly) can, however, be seen as an indication that QWG should be understood in the context of a more fundamental theory. One such theory could be the $d=4$ twistor string~\cite{Witten:2003nn,Berkovits:2004hg}, where superconformal symmetry in spacetime is explicitly preserved contrary to ordinary string theory. In fact, it was found in Ref.~\cite{Berkovits:2004jj} that its spectrum includes to the one of the $\mathcal{N}=4$ superconformal gravity, which is made up of   2 $\mathcal{N}=4$ graviton multiplets and 4 gravitini supermultiplets. One of the graviton multiplets is in the non-standard ghost-like sector. The gravitini supermultiplets contain the $15$ gauge bosons of the local $SU(4)_R$ symmetry. The $SU(4)_R$ symmetry is potentially anomalous because its coupling is chiral and, as it is gauged, its cancellation requires the dimension of the gauge group $G$ be 4. So one gets the conclusion that $G=SU(2)\times U(1)$ or $U(1)^4$. An analogous condition is consistent with, but not implied by worldsheet arguments based on the worldhsheet conformal anomaly. Apart form the $SU(4)_R^3$, also spacetime conformal anomaly is possible and it has actually been argued that $\mathcal{N}=4$ superconformal gravity can be made UV-finite and therefore anomaly-free by coupling it to exactly four $\mathcal{N}=4$ super Maxwell multiplets \cite{Fradkin6}. The computation was carried out at one-loop order, but the result should hold at all orders as the $\beta$-function in $\mathcal{N}>1$ CSG and conformal anomaly of super YM may receive contributions only from one loop. In the case of pure extended supergravity, it would actually be quite interesting to perform the analysis of functional RG in a situation where a nontrivial IR FP at one-loop order would be an exact result and where also the absence of an $R^2$ would be granted at any perturbative order. In the case of $\mathcal{N}=4$ one-loop exactness is also a consequence of the fact that all these anomalies are related by supersymmetry, and vanish when the $SU(4)_R^3$  anomaly does. As the latter is a chiral one-loop exact anomaly, its cancellation should entail the cancellation of conformal anomalies and other anomalies to all order. The version of $\mathcal{N}=4$ CSG that was actually found in the context of twistor-string is a ``non-minimal'' one where the 4-derivative complex scalar $\phi$ couples to the Weyl graviton through a term of the kind $f(\phi)\left(C_{\mu\nu\rho\sigma}\right)^2$. Imposing a manifest $SU(1,1)\simeq SL(2,R)$ requires $f$ be a constant so that one can recover the ``minimal'' version of CSG. It is quite interesting that only recently explicit actions for these theories have been constructed~\cite{Buchbinder:2012uh,Ciceri:2015qpa,Butter:2016mtk} making a RG flow analysis viable.  The anomaly-free theory described above could provide a non-Gaussian UV fixed point for this analysis. We also would like to point out that the role of superconformal symmetry in cosmological applications has found large resonance in recent years~\cite{Kallosh:2000ve,Ferrara:2010in} and in this context the relation between higher derivative terms and emergent order-parameter fields as discussed in subsection~\ref{5.1.a} has been also highlighted (cf., e.g., Ref.~\cite{Cecotti:2014ipa})

In passing, we note that trace anomalies have also been recently discussed as valuable instruments to follow the violation of conformal symmetry along the RG flow. In particular in four dimensions any such flow can be reinterpreted in terms of a spontaneously broken conformal symmetry. This has led to a deeper understanding of the a-theorem~\cite{Schwimmer:2010za,Komargodski:2011vj} and it would be definitely interesting to understand also the RG flow of QWG in this framework.

Apart from these more conceptual issues, we also discussed a cosmology that is implied by the broken phase of QWG. In the one-loop approaximation we have been able to map the broken-phase effective action on a two-field hybrid inflationary model that in its low-energy phase approaches the Starobinsky $f(R)$ model with a gravi-cosmological constant.
In particular, the inflationary potential obtained contains two scalar fields that  interact  via  derivative  coupling: scalaron $\psi$ and Hubbard--Stratonovich order-parameter field $\sigma$ (basically dilaton).  The scalaron appears when we transform the broken-phase effective action to the Einstein frame and due to its slow-role potential it plays the role of an inflaton.  The  Hubbard--Stratonovich scalar, on the other hand, results from the dynamical transmutation od the spurion HS field and it mediates the inflationary potential.
The derivative interaction between the two scalars provides a viable mechanism for a reheating scenario and graceful exit.
Requirement that Einstein's $R$ term in the low energy actions must have a coupling constant $1/2\kappa^2$
ties up the values of Starobinsky's inflation parameter $\xi$ and the gravi-cosmological constant $\Lambda$. This in turn fixes the symmetry-breakdown scale for QWG to be at about the GUT inflationary scale. Moreover, the existence of a regime where gravity is approximately scale invariant (fixed-point regime and departure of the RG flow from it) provides a simple and natural interpretation for the nearly-scale-invariance of the power spectrum of temperature fluctuations in the CMB.

It should be stressed to for simplicity's sake our cosmological considerations were done in flat Minkowski background. In order to get more refined (and also more realistic) information on the inflationary potential we should perform explicit computations of the broken-phase effective potential, in a non-trivial but cosmologically pertinent background, namely on MSS. These backgrounds are tailor-made for treatment of inflation and, in addition, they can provide us with some further guidance for the possible resolution of the cosmological constant problem on de Sitter spacetime. Work along those line is presently in progress.

\vspace{6pt}



\authorcontributions{All authors jointly discussed, conceived and wrote the manuscript.}

\funding{P.J.  was  supported  by the Czech  Science  Foundation Grant No. 19-16066S. J.K. was supported by the Grant Agency of the Czech Technical University in Prague, Grant No. SGS19/183/OHK4/3T/14.
S.G. was supported by the Israel Science Foundation (ISF), grant No. 244/17.}

\acknowledgments{Authors are grateful to Leonardo Modesto and Philip Mannheim for fruitful discussions.
}

\conflictsofinterest{The authors declare no conflict of interest.}
\abbreviations{The following abbreviations are used in this manuscript:\\

\noindent
\begin{tabular}{@{}ll}
FNSPE & Faculty of Nuclear Sciences and Physical Engineering \\
WG & Weyl gravity \\
YM & Yang--Mills\\
QWG & Quantum Weyl gravity \\
RG & Renormalization group\\
FRG & Functional renormalization group\\
FP & Fixed point\\
TP & Turning point\\
GB & Gauss--Bonnet\\
HS & Hubbard--Stratonovich\\
UV & ultraviolet\\
IR & infrared\\
MSS & Maximally Symmetric Spaces\\
CSG & Conformal Supergravity\\
SYM & Super-Yang--Mills\\
HDG & Higher derivative gravity\\
QG & Quantum gravity\\
CA & conformal anomaly\\
AS & asymptotic safety
\end{tabular}
\begin{tabular}{@{}ll}
CFT & Conformal field theory\\
SM & Starobinsky model\\
QFT & Quantum Field Theory\\
RHS & right hand side\\
LHS & left hand side\\
QCD & Quantum Chromodynamics\\
QED & Quantum Electrodynamics\\
AF & Asymptotic Freedom\\
AdS & anti-de Sitter\\
EOM & equation of motion\\
DIMREG & dimensional regularization\\
ODE & ordinary differential equation\\
FT & Fradkin--Tseytlin\\
GR & General Relativity\\
SSB & spontaneous symmetry breaking\\
FRW & Friedmann--Robertson--Walker\\
CMB & cosmic microwave background\\
VEV & vacuum expectation value\\
GUT & grand unified theory
\end{tabular}}

%

\reftitle{References}


\begin{thebibliography}{999}
\bibitem{YM} Yang,~C.N., Mills,~R., {\em Phys. Rev.} {\bf 1954}, {\em 96}, 191.

\bibitem{Utiyama} Utiyama,~R., {\em Phys. Rev.} {\bf 1956}, {\em 101}, 1597.

\bibitem{Kibble} Kibble,~T.W.B., {\em J. Math. Phys.} {\bf 1961}, {\em 2}, 212.
\bibitem{Neeman} N\'{e}eman,~Y., Regge,~T., {\em Nuovo Cimento} {\bf 1971}, {\em 1}, 1.
\bibitem{Ivanov} Ivanov,~E.A., Niederle,~J., {\em Phys. Rev. D} {\bf 1982}, {\em 25}, 976.

\bibitem{Maldacena:1997re} Maldacena,~J.M.,
{\em Int. J. Theor. Phys.} {\bf 1999} {\em 38}, 1113.

\bibitem{Borsten:2020bgv} Borsten,~L.,
{\em Riv. Nuovo Cim.} {\bf 2020}, {\em 43}, 97.

\bibitem{Taubes1} Taubes,~C.H.,
{\em Inventiones Math.} {\bf 1988}, {\em 94} 327.
\bibitem{Taubes2} Taubes,~C.H., {\em Jour. Differential Geometry} {\bf 1982}, {\em 17}, 139.


\bibitem{Bach1} Bach,~R., {\em Math. Zeitschr.} {\bf 9}, 110, (1921).

\bibitem{Kaku} Kaku,~M., Townsend,~P., Nieuwenhuizen,~P.V., {\em Phys. Lett. B} {\bf 1977}, {\em 69}, 304.

\bibitem{Strominger} Strominger,~A., Horowitz,~G.T., Perry,~M.J., {\em Nucl. Phys. B} {\bf 1984}, {\em 238}, 653.
\bibitem{Hartnoll} Hartnoll,~S.A., Policastro,~G., {\em Adv. Theor. Math. Phys.} {\bf 2006}, {\em 10}, 181.

\bibitem{Hill} Hill,~T.C., arXiv:hep-ph/0510177.

\bibitem{Maggiore} Maggiore,~M., arXiv:1506.06217.

\bibitem{JLK} Jizba, P., Rachwa{\l}, L., K\v{n}ap, J., {\em Phys. Rev. D} {\bf 2020}, {\em 101}, 044050.

\bibitem{Percacci1} Codello,~A., Percacci,~R., {\em Phys. Rev. Let.} {\bf 2006}, 97, 221301.

\bibitem{Cornwall} Cornwall,~J.M., {\em Phys. Rev. D} {\bf 1982}, 26, 1453.

\bibitem{Holdom} Holdom,~B., Ren,~R., {\em Phys. Rev. D} {\bf 2016}, 93, 124030.


\bibitem{shapiro2}
Asorey,~M., Gorbar,~E.V., Shapiro,~I.L., 
 {\em Class. Quant. Grav.} {\bf 2003}, {\em 21},  163.

\bibitem{BD} Boulware,~D.G., Deser,~S.,  {\em Phys.\ Rev.\ D} {\bf 1972}, {\em 6}, 3368.


\bibitem{Lee} Lee,~T.D., Wick,~G.C.,
  {\em Nucl.\ Phys.\ B} {\bf 1969}, {\em 9}, 209;
  {\em Phys.\ Rev.\ D} {\bf 1970}, {\em 2}, 1033.

\bibitem{Anselmi} Anselmi,~D.,  {\em JHEP} {\bf 2018}, {\em 1802}, 141;   {\em Class.\ Q.\ Grav.} {\bf 2019}, {\em 36}, 065010; {\em JHEP} {\bf 2018}, {\em 1811}, 021.

\bibitem{Tkach} Tkach,~V.I.,
  {\em Mod.\ Phys.\ Lett.\ A} {\bf 2012}, {\bf 27}, 1250131.

\bibitem{Smilga} Smilga,~A.V.,
  {\em J.\ Phys.\ A} {\bf 2014} {\em 47}, 052001.


\bibitem{Tomboulis} Tomboulis,~E.,
  {\em Phys.\ Lett.} {\bf 1977}  {\em 70B}, 361.

\bibitem{Kaku2}
  Kaku,~M.,
  {\em Phys.\ Rev.\ D} {\bf 1983},  {\em 27}, 2819.



\bibitem{Shapiro}
  Cusin,~G., de O.Salles,~F., Shapiro,~I.L.,
  {\em Phys.\ Rev.\ D} {\em 2016}, {\em 93}, 044039.


\bibitem{Asorey}
Asorey,~M., Rachwal,~L., Shapiro,~I.,
{\em Galaxies} {\bf 2018}, {\em 6}, 23.

\bibitem{Donoghue}
Donoghue,~J.F., Menezes,~G.,
{\em Phys.\ Rev.\ D} {\bf 2019}, {\em 100}, 105006.


\bibitem{Bender}   Bender,~C.M., Mannheim,~P.D.,
  {\em Phys.\ Rev.\ Lett.} {\bf 2008},  {\em 100}, 110402;
  {\em Phys.\ Rev.\ D} {\bf 2008}, {\em 78}, 025022.

\bibitem{Hartle} Hartle,~J.B., {\em Phys. Rev. D} {\bf 1994}, {\em 49}, 6543.

\bibitem{Politzer} Politzer,~H.D., {\em Phys. Rev. D} {\bf 1992}, {\em 46}, 4470.

\bibitem{Lloyd} Lloyd,~S., Maccone,~L., Garcia-Patron,~R., Giovannetti,~V., Shikano,~Y., {\em Phys. Rev. D} {\bf 2011}, {\em 84}, 025007.


\bibitem{Reuter}Reuter,~M., {\em Phys. Rev. D} {\bf 1998}, {\em 57}, 971; {\em Nucl. Phys. B} {\bf 1994}, {\em 427}, 291; {\em Nucl. Phys. B} {\bf 1994}, {\em 417}, 181.

\bibitem{Wetterich} Wetterich.~C., {\em Phys. Lett. B} {\bf 1993}, {\em 301}, 90.

\bibitem{codello} Codello, A.,  Percacci, R., Rachwal, L.,  Tonero, A., {\em  Eur.\ Phys.\ J.\ C} {\bf 2016} {\em 76}, no. 4, 226.


\bibitem{Percacci_R} Percacci,~R., {\em An Introduction to Covariant QuantumGravity and Asymptotic Safety} (World Scientific, NewYork, 2017).


\bibitem{FradkinT1} Fradkin,~E.S., Tseytlin,~A.A.,  {\em Phys.\ Rept.} {\bf 1985}, {\em 119}, 233.


\bibitem{Duff} Capper,~D.M., Duff,~M.J., {\em Phys. Lett. A} {\bf 1975}, {\em 53}, 361.


\bibitem{Berkovits:2004jj} Berkovits,~N., Witten,~E.,
{\em JHEP} {\bf 2004}, {\em 08}, 009.

\bibitem{Polchinski} Polchinski,~J., {\em String Theory}, 1st
\& 2nd Vol, (Cambridge University Press, Cambridge, 1998).



\bibitem{Weyl1} Weyl,~H., {\em Math. Zeitschr.} {\bf 2018}, {\em 2}, 384.


\bibitem{Mannheim:2017}  e.g., Mannheim,~P.D., {\em Prog.\ Part.\ Nucl.\ Phys.} {\bf 2017}, {\em 94}, 125.


\bibitem{Nakahara} Nakahara,~M., {\em Geometry, Topology and Physics}, (IOP Publishing, Ltd. Bristol, 1990).

\bibitem{confreview} Rachwal, L.,
{\em Universe} {\bf 2018} {\em 4}, 125.


\bibitem{spcompl0} Modesto, L., Rachwal, L.,
{\em J.\ Phys.\ Conf.\ Ser.} {\bf 2017}  {\em 942}, 012015.


\bibitem{finconfqg} Modesto,~L., Rachwal,~L.,
arXiv:1605.04173 [hep-th].

\bibitem{Tseytlin} Fradkin,~E.S., Tseytlin,~A.A., {\em Phys.~Rep.} {\bf 1985},~{\em 119},  233.


\bibitem{Hamber} Hamber,~H.W., {\em Quantum Gravitation, The Feynman Path Integral Approach}, (Springer, Berlin, 2009).

\bibitem{DeWitt1} DeWitt,~B.S., {\em General Relativity:  An Einstein Centenary Survey}, Hawking,~S.W., and Israel,~W., editors, (Cambridge Univ. Press, Cambridge, 1979).

\bibitem{DeWitt2} DeWitt,~B.S., {\em Relativity, Groups and Topology II}, DeWitt,~B.S., and Stora,~R. editors, (North-Holland, Amsterdam, 1984).

\bibitem{Carlip} Carlip,~S., {\em Class. Q. Grav.} {\bf 1998}, {\em 15}, 2629.

\bibitem{Freedman} Freedman,~M.H., {\em J. Differential Geom.} {\bf 1998}, {\em 17}, 357.



\bibitem{Irakleidou} Irakleidou.~M., Lovrekovic,~I., {\em Phys. Rev. D} {\bf 2016}, {\em 93}, 104043.


\bibitem{Wetterich2} Wetterich,~C., {\em Phys. Rev. D} {\bf 2018}, {\em 98}, 026028.


\bibitem{supplement} See Supplemental Material at
\href{http://link.aps.org/supplemental/10.1103/PhysRevD.101.044050}{http://link.aps.org/supplemental/10.1103/}\\
\href{http://link.aps.org/supplemental/10.1103/PhysRevD.101.044050}{PhysRevD.101.044050} for finer
technical details.


\bibitem{Litim2}Litim,~D.F.,
  {\em Phys.\ Rev.\ D} {\bf 2001}, {em 64}, 105007;  {\em Phys.\ Lett.\ B} {\bf 2000}, {\em 486}, 92.


\bibitem{Fradkin6}
  E.S.~Fradkin,~E.S., Tseytlin,~A.A.,
  {\em Phys.\ Lett. B} {\bf 1984}, {\em 134B}, 187.


\bibitem{thooft} 't~Hooft,~G., {\em Under the spell of the gauge principle}, (World Scientific Publishing, Singapore, 1994).

\bibitem{Bonora:1985cq}
Bonora~L., Pasti,~P., M.~Bregola,~M.,
{\em Class. Quant. Grav.} {\bf 1986}, {\em 3}, 635.

\bibitem{Christensen:1978gi}
S.~Christensen and M.~Duff,
{\em Phys. Lett. B }{\bf 1978} {\em76},  571

\bibitem{Christensen:1978md}
S.~Christensen and M.~Duff,
{\em Nucl. Phys. B }{\bf 1979} {\em154}, 301-342



\bibitem{Weinberg} Weinberg,~S., {\em Ultraviolet divergences in quantum theories of gravitation} in {\em General Relativity: An Einstein Centenary Survey},
eds Hawking~S.W., Israel,W., (Cambridge University Press, Cambridge, England, 1979), Chapter 16,  pp.790--831.





\bibitem{expjizba}Jizba,~P., Kleinert,~H., Scardigli,~F.,  {\em Eur.\ Phys.\ J.\ C} {\bf 2015}, {\em 75}, 245.






\bibitem{Hubbard} Hubbard,~K., {\em Phys. Rev. Lett.}  {\bf 1959}, {\em 3}, 77.

\bibitem{Stratonovich} Stratonovich,~R.L., {\em Soviet Physics Doklady} {\bf 1958}, {\em 2}, 416.


\bibitem{GL} Ginzburg,~V.L., Landau,~L.D., {\em Zh. Eksp. Teor. Fiz.} {\bf 1950}, {\em 20}, 1064.


\bibitem{Stevenson} Stevenson,~P.M. {\em Phys. Rev. D} {\bf 1981}, {\em 23}, 2916.

\bibitem{Kleinert-QFT} Kleinert,~H., Schulte-Frohlinde,~V., {\it Critical Properties of $\phi^4$-Theories} (World Scientific, Singapore, 2001).


\bibitem{Fulling:74}
S.~Fulling, L.~Parker, and B.~Hu,
Phys.\ Rev.\ D {\bf 10},  3905 (1974).


\bibitem{exactsol} Li~Y.D., Modesto~L., Rachwal,~L.,
  {\em J. High Energy Phys} {\bf 2015}, {\em 1512}, 173.


\bibitem{Ade}Ade,~P.A.R.,  et al., BICEP2/Keck and Planck Collaborations, {\em Phys. Rev. Lett.} {\bf 2015}, {\em 114}, 101301.
\bibitem{Planck2}    Ade,~P.A.R., et al., The Keck Array and BICEP2 Collaborations, {\em Phys. Rev. D} {\bf 2017}, {\em 96}, 102003.
\bibitem{Mannheim:2011is}
Mannheim,~P.D.,
{\em Phys.\ Rev.\ D} {\bf 2012}, {\em 85}, 124008.

\bibitem{Sachdev} Sachdev,~S., {\em Quantum Phase Transitions}, (Cambridge  University Press, Cambridge, 2011).
\bibitem{AS} Altland,~A., Simons,~B., {\em Condensed Matter Field Theory}, (Cambridge University Press, Cambridge, 2013).
\bibitem{MS1}Mathews,~P.T., Salam,~A., {\em Nuovo Cimento} {\bf 1954},~{\em 12}, 563.
\bibitem{MS2}Mathews,~P.T., Salam,~A., {\em Nuovo Cimento} {\bf 1955}, {\em 2}, 120.
\bibitem{Col} Coleman,~S., {\em Aspects of Symmetry: Selected Erice Lectures}, (Cambridge University Press, Cambridge, 1988).
\bibitem{KL3} Kleinert,~H., {\em On the hadronization of quark theories}, in Proceedingsof the Erice Summer Institute 1976, {\em Understanding the Fundamental Constituents of Matter}, ed. by Zichichi,~A., (Plenum Press, NewYork, 1978), pp. 289–390.

\bibitem{Stev} Stevenson,~P.M., {\em Phys. Rev. D} {\bf 1981}, {\em 23}, 2916.
\bibitem{Bender2} Bender,~C.M.,   Pinsky,~K.S.,  Simmons,~L.M., {\em J. Math. Phys.} {\bf 1999}, {\em 30}, 1447.
\bibitem{KS-F} Kleinert,~H., Schulte-Frohlinde,~V.,{\em Critical Propertiesof $\varphi^4$-Theories}, (World Scientific, Singapore, 2001).
\bibitem{knap} Urban,~P., {\em MSc thesis, FNSPE, Czech Technical University in Prague}, {\bf 2018}, \\
\href{https://physics.fjfi.cvut.cz/publications/mf/2018/dp_mf_18_urban.pdf}{https://physics.fjfi.cvut.cz/publications/}
\href{https://physics.fjfi.cvut.cz/publications/mf/2018/dp_mf_18_urban.pdf}{mf/2018/urban.pdf}



\bibitem{Birrel} Birrell,~N.D., Davies,~P.C.W., {\em Quantum Fields in Curved Space}, (Cambridge University Press, Cambridge, 1982).

\bibitem{Whitt} Whitt,~B., {\em Phys. Lett. B} {\bf 1984}, {\em 145}, 176.
\bibitem{Capozziello} Capozziello,~S., Faraoni,~V., {\em Beyond Einstein Gravity; A Survey of Gravitational Theories for Cosmology and Astrophysics}, (Springer, London, 2011).


\bibitem{Planck3} Akrami,~Y., et al., [Planck, Planck 2018 results. X. Constraints on inflation],
[arXiv:1807.06211 [astro-ph.CO]].

\bibitem{Linde} Kallosh,~R., Linde,~A., {\em JCAP} {\bf 2013},~{\em 07}, 002.
\bibitem{Costa} Costa,~R., H.~Nastase,~H., {\em JHEP} {\bf 2014}, {\em 06}, 145.
\bibitem{Turok} Bars,~I., Steinhardt,~P., Turok,~N., {\em Phys. Rev. D} {\bf 2014},~{\em 89}, 043515.
\bibitem{Vanzo1} Rinaldi,~M., Cognola,~G., Vanzo,~L., Zerbini,~S., {\em JCAP}~{\bf 2014}, {\em 1408}, 015.
\bibitem{Schwimmer:2010za}
Schwimmer,~A., Theisen,~S.,
{\em Nucl. Phys. B} {\bf 2011}, {\em 847}, 590.

\bibitem{Fayet:1978ig}
Fayet,~P.,
{\em Nucl. Phys. B} {\bf 1979}, {\em 149}, 137.

\bibitem{hooftanomaly}
’t Hooft, G., {\em NATO Adv. Study Inst. Ser. B Phys.} {\bf 1980}, {\em 59}, 135.

\bibitem{Berkovits:2004hg} Berkovits,~N.,
{\em Phys. Rev. Lett.} {\bf 2004}, {\em 93}, 011601.


\bibitem{Witten:2003nn} Witten,~E.,
{\em Commun. Math. Phys.} {\bf 2004}, {\em 252}, 189.
Jun 2020

\bibitem{Buchbinder:2012uh} Buchbinder,~I., Pletnev,~N., Tseytlin,~A.,
{\em Phys. Lett. B} {\bf 2012}, {\em 717}, 274.

\bibitem{Ciceri:2015qpa} Ciceri,~F., Sahoo,~B.,
{\em JHEP} {\bf 2016}, {\em 01}, 059.

\bibitem{Butter:2016mtk} Butter,~D., Ciceri,~F., de Wit,~B., Sahoo,~B.,
{\em Phys. Rev. Lett.} {\bf 2017}. {\em 118}, 081602.

\bibitem{Kallosh:2000ve} Kallosh,~R., Kofman,~L., Linde,~A.D., Van Proeyen,~A.,
{\em Class. Quant. Grav.} {\bf 2000}, {\em 17}, 4269.

\bibitem{Ferrara:2010in}
Ferrara,~S., Kallosh,~R., Linde,~A.D., Marrani,~A., Van Proeyen,~A.,
{\em Phys. Rev. D} {\bf 2011}, {\em 83}, 025008.

\bibitem{Cecotti:2014ipa} Cecotti,~S., Kallosh,~R.,
{\em JHEP} {\bf 2014}, {\em 05}, 114.

\bibitem{Schwimmer:2010za}
Schwimmer, A., Theisen, S.,
{\em Nucl. Phys. B} {\bf 2011} {\em 847}, 590.

\bibitem{Komargodski:2011vj}
Komargodski, Z., Schwimmer, A.,
{\em JHEP} {\bf 2011} {\em 12}, 099.

%
%
%


\end{thebibliography}
\end{document}